%% file: EXO-17-017_temp.tex
\pdfoutput=1

\documentclass[11pt,twoside,a4paper,cmspaper,final,collab]{cms-tdr}
\def\svnVersion{480603P}\def\cmsCernNoTag{CERN-EP-2018-219}\def\cmsCernDate{\today}\def\cmsMessage{Published in Physical Review D as \href{http://dx.doi.org/10.1103/PhysRevD.98.092001}{\doi{10.1103/PhysRevD.98.092001.}}}

\begin{document}\cmsNoteHeader{EXO-17-017}

\hyphenation{had-ron-i-za-tion}
\hyphenation{cal-or-i-me-ter}
\hyphenation{de-vices}
\RCS$Revision: 480191 $
\RCS$HeadURL: svn+ssh://svn.cern.ch/reps/tdr2/papers/EXO-17-017/trunk/EXO-17-017.tex $
\RCS$Id: EXO-17-017.tex 480191 2018-11-02 21:44:09Z abuccill $

\newlength\cmsFigWidth
\ifthenelse{\boolean{cms@external}}{\setlength\cmsFigWidth{0.98\columnwidth}}{\setlength\cmsFigWidth{0.6\textwidth}}

\providecommand{\cmsTable}[1]{\resizebox{\textwidth}{!}{#1}}
\newlength\cmsTabSkip\setlength{\cmsTabSkip}{1ex}

\ifthenelse{\boolean{cms@external}}{\providecommand{\CL}{C.L.\xspace}}{\providecommand{\CL}{CL\xspace}}

\newcommand{\sieie}{\ensuremath{\sigma_{\eta \eta}}\xspace}
\newcommand{\mgg}{\ensuremath{m_{\gamma\gamma}}\xspace}
\newcommand{\MS}{\ensuremath{M_{\mathrm{S}}}\xspace}
\newcommand{\nED}{\ensuremath{n_{\mathrm{ED}}}\xspace}
\newcommand{\rc}{\ensuremath{r_{\mathrm{c}}}\xspace}
\newcommand{\GammaX}{\ensuremath{\Gamma_{\mathrm{X}}}\xspace}
\newcommand{\mX}{\ensuremath{m_{\mathrm{X}}}\xspace}
\newcommand{\mS}{\ensuremath{m_{\mathrm{S}}}\xspace}
\providecommand{\GammaXOmX}{\ensuremath{\GammaX/ \mX}\xspace}
\providecommand{\Zee}{\ensuremath{\cPZ \to \Pep \Pem}\xspace}
\providecommand{\ktild}{\ensuremath{\tilde{k}}\xspace}
\providecommand{\Grav}{\PXXG\xspace}
\providecommand{\mG}{\ensuremath{m_\mathrm{\Grav}}\xspace}

\cmsNoteHeader{EXO-17-017}

\title{Search for physics beyond the standard model in high-mass diphoton events from proton-proton collisions at \texorpdfstring{$\sqrt{s} = 13\TeV$}{sqrt(s) = 13 TeV}}

\date{\today}

\abstract{A search for physics beyond the standard model is performed using a sample of high-mass diphoton events produced in proton-proton collisions at $\sqrt{s} = 13\TeV$. The data sample was collected in 2016 with the CMS detector at the LHC and corresponds to an integrated luminosity of 35.9\fbinv. The search is performed for both resonant and nonresonant new physics signatures. At 95\% confidence level, lower limits on the mass of the first Kaluza--Klein excitation of the graviton in the Randall--Sundrum warped extra-dimensional model are determined to be in the range of 2.3 to 4.6\TeV, for values of the associated coupling parameter between 0.01 and 0.2. Lower limits on the production of scalar resonances and model-independent cross section upper limits are also provided. For the large extra-dimensional model of Arkani-Hamed, Dimopoulos, and Dvali, lower limits are set on the string mass scale \MS ranging from 5.6 to 9.7\TeV, depending on the model parameters. The first exclusion limits are set in the two-dimensional parameter space of a continuum clockwork model.}

\hypersetup{%
pdfauthor={CMS Collaboration},%
pdftitle={Search for physics beyond the standard model in high-mass diphoton events from proton-proton collisions at sqrt(s) = 13 TeV},%
pdfsubject={CMS},%
pdfkeywords={CMS, physics, diphoton, extra dimensions, RS, ADD, clockwork}}

\maketitle

\section{Introduction}\label{sec:intro}

While the standard model (SM) of particle physics has been an enormously successful description of observed phenomena, it is still widely believed to be incomplete. In the SM, the Higgs boson mass receives quantum corrections from loops containing SM particles. Because the Higgs boson is a fundamental scalar, the magnitude of the mass corrections is set by the cutoff parameter of the loop integrals and the only natural mass scale in the SM that can act as a cutoff is the Planck scale ($\Mpl \sim 10^{19}\GeV$) at which quantum gravity is expected to emerge. Therefore, unless the Higgs boson mass parameter is fine-tuned to an extreme degree, there must exist some new physics beyond the SM to constrain these quantum corrections and stabilize the mass of the Higgs boson. Many models for such new physics have been proposed. We consider three such models in this paper.

Through their modification of the effective Planck scale, extra spatial dimensions have been proposed as a possible solution to this hierarchy problem~\cite{hp1,hp2}, which arises from the large difference between the gravitational and electroweak scales. In the model proposed by Arkani-Hamed, Dimopoulos, and Dvali (ADD)~\cite{ArkaniHamed:1998rs,Antoniadis:1998ig,ArkaniHamed:1998nn}, the existence of \nED additional spatial dimensions, compactified with an average radius \rc, produces an effective Planck mass $M_{\mathrm{Pl}}$ in our four-dimensional (4D) world that is related to the true Planck mass by $M^2_{\mathrm{Pl}} \sim M^{2+\nED}_{\mathrm{Pl}(4+\nED)} \rc^{\nED}$. It is therefore possible, with appropriate values for \nED and \rc, that the value of the Planck scale in (4+\nED)-dimensional spacetime, $M_{\mathrm{Pl}(4+\nED)}$, could be of the order of the electroweak scale, thus solving the SM hierarchy problem, while still producing the much larger apparent Planck scale that we observe in our 4D world. In effect, the true strength of gravity could actually be comparable to the electroweak force, and it merely appears weaker because of its propagation in the extra dimensions.  At the same time, the SM gauge forces and particles are confined to our 4D spacetime.

Randall and Sundrum (RS) proposed an alternative model~\cite{Randall:1999ee,Randall:1999vf} with just one additional dimension that has a warped geometry, described by a curvature parameter $k$. The extra dimension is compactified with radius \rc.  From the point of view of a 4D observer, a fundamental mass parameter $m_0$ defined on the SM brane in the full five-dimensional (5D) theory will appear as a physical mass $m=\re^{-k\rc\pi}m_0$. In this model, solving the hierarchy problem requires that the fundamental mass should be $m_0 \sim \Mpl$ and the observed mass should be at the {\TeVns} scale.  Because of the exponential warp factor, this large hierarchy can be generated if $k\rc \approx 11$--12, thus requiring very little fine tuning.

A third proposed solution to the hierarchy problem that we consider in this paper is the continuum clockwork mechanism~\cite{Giudice2017}, which coincides with a 5D gravitational theory on a linear dilaton background~\cite{Antoniadis:2001sw,Antoniadis:2011qw}.  The clockwork~\cite{Choi:2015fiu,Kaplan:2015fuy} is a general mechanism that can introduce large effective interaction scales from dynamics occurring at much lower energies.  This is achieved by introducing $N$ copies of some particle content on different sites forming a one-dimensional lattice in theory space.  The physical mass spectrum consists of a single massless mode localized on the end site of the lattice and a set of massive modes (`gears')  distributed along the sites.  In the continuum limit of the clockwork with $N\to\infty$, this lattice is interpreted as a physical extra dimension.

In all three models, deviations from the SM expectations should be evident at the CERN LHC through Kaluza--Klein (KK) modes of the graviton, which couple to the SM through the stress-energy tensor and decay into two SM particles. Searches for pairs of high-mass photons are favorable because the SM backgrounds are lower and the mass resolution is better than in the dijet channel, and the branching fraction to diphoton final states is larger than that to dilepton final states. In the ADD model, the compactification of the extra dimensions gives rise to a KK series of virtual graviton states because the momenta along the extra dimensions are quantized and appear as additional contributions to the graviton effective mass. Here, the KK modes are very closely spaced, leading to an effective nonresonant enhancement of the diphoton spectrum at high invariant mass (\mgg). In the RS model, the presence of the additional spatial dimension quantizes the masses of the KK states. These states are widely spaced and become narrow in the limit of small $\tilde{k}\equiv k/\overline{\Mpl}$, where $\overline{\Mpl} = \Mpl/\sqrt{\smash[b]{8\pi}}$ is the reduced Planck mass. In the continuum clockwork mechanism, the gears play the role of the KK modes.  The usual massless graviton is accompanied by an infinite tower of massive spin-2 graviton gears with a characteristic pattern of masses and couplings.  In particular, the masses of the graviton gears can be so densely distributed that they produce an approximately continuous contribution to the diphoton spectrum as a function of \mgg~\cite{Baryakhtar:2012wj,Giudice:2017fmj}, much like in the ADD model. However, unlike the ADD model, the KK modes are entirely on shell, as in the RS case, so interference effects are negligible.

In addition to the above models that could address the hierarchy problem, high-mass diphoton events are also potentially sensitive to other beyond-SM physics, such as the decays of heavy spin-0 resonances. These spin-0 resonances could arise from extended Higgs sectors~\cite{Branco:2011iw,PhysRevD.8.1226,Craig:2013hca}. Model-independent cross section limits can be obtained on generic production of exotic spin-0 and spin-2 resonances decaying to pairs of photons.

Searches for new physics in the high-mass diphoton channel from Run 2 of the LHC were previously performed by the ATLAS and CMS experiments using $\Pp\Pp$ collisions at a center-of-mass energy $\sqrt{s}=13\TeV$~\cite{Aaboud:2017yyg,Khachatryan:2016yec,Aaboud:2016tru,Khachatryan:2016hje}. Prior searches have also been performed by both experiments in Run 1 of the LHC~\cite{Aad:2015mna,Khachatryan:2015qba,Aad:2012cy,Chatrchyan:2011fq}, at $\sqrt{s}=7$ and 8\TeV, and also at the Tevatron by the CDF~\cite{Aaltonen:2011xp} and \DZERO~\cite{Abazov:2010xh} experiments using $\Pp\Pap$ collisions at $\sqrt{s}=1.96\TeV$.

We present new results from a search for beyond-SM physics in the high-mass diphoton spectrum, using data collected with the CMS detector in 2016 corresponding to an integrated luminosity of 35.9\fbinv. Two complementary background estimation techniques are used. For a search for resonant excesses such as in the RS model, we implement a technique where the diphoton spectrum is fit to a parametrized functional form, allowing for a fully data-driven description of the shape. To search for nonresonant deviations from the SM prediction like those that arise in the ADD and clockwork models, a next-to-next-to-leading-order (NNLO) calculation of the SM diphoton background is performed and the background from jets being misidentified as photons is estimated from control samples in data.

The remainder of this document is organized as follows. The CMS detector is described in Section~\ref{sec:cms}. Event reconstruction and selection are outlined in Section~\ref{sec:event_selection}. Section~\ref{sec:signals} describes the simulation of extra-dimensional signal models, and the background determinations for the searches for resonant and nonresonant excesses are described in Section~\ref{sec:background}. Sources of systematic uncertainty are discussed in Section~\ref{sec:systematics}. We present our results in Section~\ref{sec:results}.

\section{The CMS detector}\label{sec:cms}

The CMS detector is a multi-purpose collider detector at the LHC.
The central feature of the CMS apparatus is a superconducting solenoid of 6\unit{m} internal diameter, providing a magnetic field of 3.8\unit{T}. Within the solenoid volume are a silicon pixel and strip tracker, a lead tungstate crystal electromagnetic calorimeter (ECAL), and a brass and scintillator hadron calorimeter (HCAL), each composed of a barrel and two endcap sections. Forward calorimeters extend the pseudorapidity ($\eta$) coverage provided by the barrel and endcap detectors. Muons are detected in gas-ionization chambers embedded in the steel flux-return yoke outside the solenoid.

The ECAL barrel (EB) provides coverage in the range $\abs{\eta} < 1.48$. This is extended by each ECAL endcap (EE) to $1.48 < \abs{\eta} < 3$. In the EB, photons that have energies above the range of tens of GeV have an energy resolution of about 1.3\% up to $\abs{\eta} = 1$, rising to about 2.5\% at $\abs{\eta} = 1.4$. In each EE, photons in this energy range have a resolution between 3 and 4\%~\cite{Khachatryan:2015iwa}.

Events of interest are selected using a two-tiered trigger system~\cite{Khachatryan:2016bia}. The first level, composed of custom hardware processors, uses information from the calorimeters and muon detectors to select events at a rate of around 100\unit{kHz} within a time interval of less than 4\mus. The second level, known as the high-level trigger, consists of a farm of processors running a version of the full event reconstruction software optimized for fast processing, and reduces the event rate to around 1\unit{kHz} before data storage.

A more detailed description of the CMS detector, together with a definition of the coordinate system used and the relevant kinematic variables, can be found in Ref.~\cite{Chatrchyan:2008zzk}.

\section{Event reconstruction and selection}\label{sec:event_selection}

Individual particles in the CMS detector are reconstructed using the particle-flow event algorithm~\cite{CMS-PRF-14-001}. Photon candidates are reconstructed from energy deposits in the ECAL. Individual energy deposits are grouped into superclusters~\cite{Khachatryan:2015iwa} that are compatible with the expected shower shape extending along the azimuthal ($\phi$) direction. This allows for the recovery of the energy deposited by bremsstrahlung and photon conversions. The clustering algorithm does not make any hypothesis as to whether the particle originating from the interaction point is a photon or an electron. Thus the same algorithm used for photon reconstruction can be applied to $\Zee$ events and these events are used to measure the efficiency of the photon selection criteria and of the photon energy scale and resolution. A more detailed description of photon reconstruction in the CMS detector can be found in Ref.~\cite{Khachatryan:2015iwa}.

{\tolerance=800 Reconstructed photon candidates must pass additional identification criteria to suppress misidentified jets and electrons, while maintaining high efficiency. These criteria are based on observables sensitive to the electromagnetic shower shape and the extra activity surrounding the shower. The electromagnetic shower shape is measured using $\sigma_{\eta \eta}$, the spatial second-order moment of the photon candidate in the $\eta$ direction~\cite{Khachatryan:2015iwa}. Isolation variables are based on the total transverse momentum of particle candidates with coordinates $(\eta,\,\phi)$ reconstructed within a cone of size $\DR = 0.3$ around the photon candidate with coordinates $(\eta_\gamma,\,\phi_\gamma)$, where $\DR = \sqrt{\smash[b]{(\eta - \eta_\gamma)^2 +(\phi-\phi_\gamma)^2}}$. Separate isolation variables are defined for charged hadron and photon candidates. Electrons are vetoed based on hits in the silicon pixel and strip trackers with a further check to ensure photon candidates associated with electron tracks are incompatible with those resulting from photon conversions.\par}

Events are selected by a trigger that requires at least two reconstructed photon candidates, each with transverse momentum $\pt > 60\GeV$. For these events, the ratio of the energy deposited in the HCAL behind each photon candidate and the photon energy in the ECAL is required to be less than 0.15. To avoid non-uniform efficiency near the trigger threshold, a more stringent selection of photon $\pt > 75\GeV$ is applied offline. Events are required to have one photon in the EB with $\abs{\eta} < 1.44$, and another in either the EB or in an EE, where it must have $1.57 < \abs{\eta} < 2.5$. Events with both photon candidates in the EEs are dominated by SM production and have negligible sensitivity to the target beyond-SM signals, and thus are omitted from this analysis. Two signal regions are considered: one with both photons in the EB, denoted EBEB, and the other with one photon in the EB and the other in either EE, denoted EBEE. The invariant mass of the photon pair must satisfy a minimum requirement of $\mgg>230$ and 330\GeV in the EBEB and EBEE categories, respectively. This threshold avoids sculpting of the distribution while maintaining full efficiency in each region. In the search for nonresonant signals, this requirement is increased to $\mgg>500\GeV$ in both categories and photon pairs must additionally satisfy $\DR> 0.45$, to be consistent with the background calculation for SM diphoton production, as described in Section~\ref{sec:bkg_nonresonant}.

\begin{figure}[!htb]
    \centering
    \includegraphics[width=\cmsFigWidth]{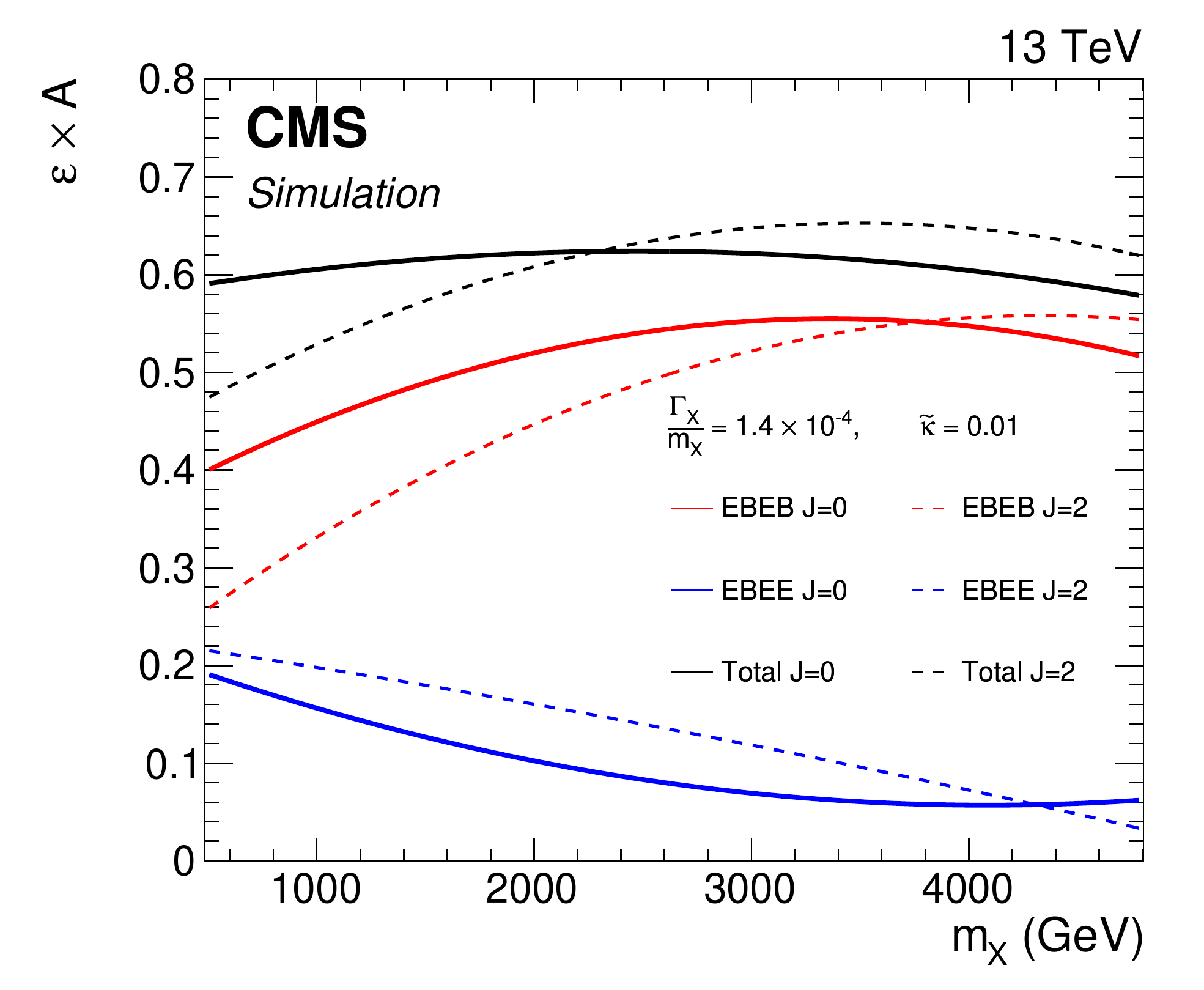}
    \caption{
      The product of the event selection efficiency ($\varepsilon$) and the detector acceptance (A) is shown as a function of signal resonance mass \mX for the $\GammaXOmX=1.4\times10^{-4}$ signal width hypothesis. The total (black), EBEB (red), and EBEE (blue) curves are shown for the spin (J) hypotheses $\mathrm{J}=0$ (solid) and $\mathrm{J}=2$ (dashed).
    }
    \label{fig:eff_x_acc}
\end{figure}

The trigger and identification efficiencies are found to be compatible within uncertainties (3\%) between data and simulation and the overall efficiency is about 90 (87)\% for single photons in the barrel (endcaps). The selection criteria as well as the level of agreement between data and simulation were determined using simulated signal, background, and control data samples and were fixed before inspecting the diphoton invariant mass distribution in the search regions. The product of the event selection efficiency ($\varepsilon$) and the detector acceptance ($A$) is shown in Fig.~\ref{fig:eff_x_acc} for two spin hypotheses in the narrowest resonance-width scenario considered in this analysis, as discussed further in Section~\ref{sec:signals}. (For the other two resonance-width hypotheses considered, the acceptance values are very close to this one.)

In order to obtain the optimal energy resolution, the ECAL signals are calibrated and corrected for several detector effects. The variation of the crystal transparency during data taking is monitored and corrected using a dedicated monitoring system and the single-channel response is equalized using collision events. The data used for this analysis were reconstructed with a detector calibration optimized for the 2016 data taking conditions. The energies of the photon candidates are first measured by the ECAL and are then corrected for shower noncontainment effects using a multivariate regression procedure. The corrections are tuned using a simulation of photon candidates with energies spanning the entire range explored by this analysis.

Differences in the photon energy scale and resolution between data and simulation are estimated using dielectron events. Energy scale and resolution corrections are derived primarily from \Zee events, with the electrons reconstructed as photons, using the procedure described in Ref.~\cite{Khachatryan:2015iwa}. The corrections are derived in eight categories defined in terms of the \RNINE\xspace variable (defined as the ratio of the energy deposited in the central $3{\times}3$ crystal matrix and the full cluster energy) and the location of the photon within the detector along the $\eta$ direction.

The size of the energy scale corrections derived from \Zee events is of the order of 0.5\%, while the additional Gaussian smearing needed to match the energy resolution of simulated events with that in data varies between 0.8 and 1.5\% for photon candidates in the EB region and between 2 and 2.5\% for photon candidates in the EE regions.

The diphoton mass resolution has contributions from the measurements of the photon energies and from the resolution of the measurement between the two photons. If the $z$ position of the vertex from which the photons originate is known to within about 10\mm, then the experimental angular resolution between the photons makes a negligible contribution to the mass resolution. Correctly associating the diphoton candidate with one of the vertices reconstructed from the charged-particle tracks in the event satisfies the above requirements since the positions of these vertices are measured with far greater precision. The interaction vertex associated with the diphoton system is selected using the algorithm described in Refs.~\cite{Khachatryan:2014ira,Sirunyan:2018ouh}. Because photons do not deposit ionization energy in the tracker, the assignment of the diphoton candidate to a vertex can only be done indirectly by exploiting the properties of each reconstructed vertex. Three discriminating variables are calculated for each reconstructed vertex: the $\pt^2$ sum of the charged-particle tracks associated with the vertex, and two variables that quantify the vector and scalar balance of \pt between the diphoton system and the tracks associated with the vertex. In addition, if either photon has an associated track that has been identified as originating from a photon conversion to an electron-positron pair, the conversion information is used. These variables provide the inputs to a multivariate classifier based on a boosted decision tree used to select the reconstructed vertex of the diphoton system. For signal events with diphoton invariant masses above 500\GeV, the fraction of events in which the interaction vertex is correctly assigned is approximately 90\%.

\section{Signal simulation}\label{sec:signals}

The ADD signal samples used in this analysis were produced at leading order (LO) using the Monte Carlo (MC) event generator \SHERPA~2.1.1~\cite{Gleisberg:2008ta} with the CT10 set of parton distribution functions (PDFs)~\cite{Lai:2010vv,Gao:2013xoa}. These simulations include the effect of interference between the ADD signal and the SM diphoton processes, which can be large. To be able to set limits on possible deviations from the SM, additional SM-only samples are generated identically, and the difference between these and the ADD samples therefore encompasses the combined effects of the ADD signal and the interference.

The implementation of the ADD model within \SHERPA is parametrized by the ultraviolet string cutoff scale \MS, which is related to the fundamental Planck scale and the number of extra dimensions \nED. Since the ADD model is an effective theory only valid below the cutoff scale, the generated diphoton mass spectra are truncated at the chosen value of \MS. The amplitude for a process involving virtual graviton exchange involves a sum over the KK tower of graviton mass states. This process can be represented by a higher-dimensional operator with coefficients suppressed by some mass scale~\cite{Gleisberg:2003ue}, which can be parametrized by $\eta_{\mathrm{G}} = \mathcal{F}/\MS^4$, where $\mathcal{F}$ is a dimensionless parameter for which several conventions exist in the literature. We consider the conventions by Giudice, Rattazzi, and Wells (GRW)~\cite{Giudice:1998ck}, by Han, Lykken, and Zhang (HLZ)~\cite{Han:1998sg}, and by Hewett~\cite{Hewett:1998sn}, expressed as:
\begin{linenomath}
\begin{equation}
	\label{eqn:add_f_conventions}
	\mathcal{F} =
	\begin{cases}
	1 \quad \text{(GRW)}, \\
	\log\left(\frac{\MS^2}{\hat{s}} \right) , \; \text{if} \; \nED = 2 \\
	\frac{2}{\nED - 2} , \; \text{if} \; \nED > 2 \\
	\pm\frac{2}{\pi} \quad \text{(Hewett)},
	\end{cases}
	\text{(HLZ)},
\end{equation}
\end{linenomath}
where $\sqrt{\hat{s}}$ is the center-of-mass energy of the colliding partons.

Signal model assumptions from different conventions but with the same value of $\eta_{\mathrm{G}}$ are equivalent, reducing the number of distinct scenarios allowed by Eq.~(\ref{eqn:add_f_conventions}). All possible choices of model parameters can be made equivalent to the signals produced using either the convention by GRW, HLZ assuming $\nED = 2$, or Hewett using $\mathcal{F} = -2/\pi$. Twelve model points for each choice are generated in the range $3 < \MS < 11\TeV$. For each model point, the CMS detector response is simulated using \GEANTfour~\cite{Agostinelli:2002hh} and includes the effects of multiple proton-proton collisions occurring within the same LHC bunch crossing, known as `pileup'.

No additional samples are needed to generate the clockwork signal; instead the ADD signal samples are reinterpreted to produce the clockwork prediction.  In the clockwork model, the KK modes are all on shell, so there is no interference effect, while the ADD prediction includes both a direct term and an interference term.  The GRW and negative Hewett models have opposite signs for the interference term, so the direct term can be isolated by linearly adding, with appropriate weights, the predictions assuming the GRW and negative Hewett conventions.  The direct term is then rescaled by Eq.~(\ref{eqn:clockwork_rescale}), provided by the authors of Ref.~\cite{Giudice:2017fmj}:
\begin{linenomath}
\ifthenelse{\boolean{cms@external}}{
\begin{multline}
	\label{eqn:clockwork_rescale}
	\theta(\mgg-k)\frac{30 \MS^8}{283\pi M_5^3}\sqrt{1-\frac{k^2}{\mgg^2}}\\
\times\frac{1}{\mgg^5}\Bigg[1+\frac{(5^2)(7)(17)}{(283)(2^8)}\Big(1-\frac{k}{\mgg}\Big)^9\sqrt{\frac{\mgg}{k}}\Bigg]^{-1}.
\end{multline}
}{
\begin{equation}
	\label{eqn:clockwork_rescale}
	\theta(\mgg-k)\frac{30 \MS^8}{283\pi M_5^3}\sqrt{1-\frac{k^2}{\mgg^2}}\frac{1}{\mgg^5}\Bigg[1+\frac{(5^2)(7)(17)}{(283)(2^8)}\Big(1-\frac{k}{\mgg}\Big)^9\sqrt{\frac{\mgg}{k}}\Bigg]^{-1}.
\end{equation}
}
\end{linenomath}
Here, \MS is defined in the GRW convention, $M_5$ is the fundamental scale of the gravitational interactions, and $k$ is the `clockwork spring', which, phenomenologically, controls the energy scale at which the KK modes can be excited.  To solve the hierarchy problem, $M_5$ should be close to the electroweak scale. Demanding perturbativity of the theory imposes the constraint $k<M_5$.

For resonant diphoton production, the signal distribution in \mgg is determined from the convolution of the intrinsic shape of the resonance and the ECAL detector response. The intrinsic shapes of both the spin-0 and spin-2 resonant signals were derived using the \PYTHIA~8.2~\cite{Sjostrand:2014zea} event generator with the NNPDF2.3~\cite{Ball:2012cx} set of PDFs and the CUETP8M1~\cite{Khachatryan:2015pea} underlying event tune. The spin-0 signal corresponds to a heavy SM-like Higgs boson, while the spin-2 signal corresponds to the RS graviton. Three signal width hypotheses are considered: $\GammaXOmX = 1.4\times 10^{-4}$, $1.4\times 10^{-2}$, and $5.6\times 10^{-2}$, corresponding to a width narrower than, comparable to, and wider than the detector resolution, respectively. These three width hypotheses correspond, in the case of an RS graviton, to $\ktild = 0.01$, 0.1, and 0.2, respectively. A set of signal samples was simulated excluding the detector response, forming a fine grid of mass points with 125\GeV spacing. These samples are used to measure the signal kinematic acceptance and generator-level mass shape. The resulting shapes are interpolated to intermediate mass points using a parametric description of the distribution. The detector response was determined using signal samples simulated with \GEANTfour, and includes the effects of pileup. These samples were generated assuming small intrinsic width, with additional Gaussian smearing, determined using dielectron events, applied to correct the simulated resolution to that of data. Nine equidistant mass hypotheses in the range 500--4500\GeV were  employed. The signal mass resolution, quantified through the ratio of the full width at half maximum (FWHM) of the distribution, divided by 2.35, to the peak position, is roughly 1 and 1.5\% for the EBEB and EBEE categories, respectively.

For the spin-0 and spin-2 assumptions, in order to determine the signal normalization the final selection efficiency was combined with the kinematic acceptance. The former is obtained from fully simulated samples and interpolated using a quadratic function of the resonance mass; the latter is obtained from the finely spaced grid of samples and parametrized as a quadratic function of both the resonance mass and its width.

\section{Background determination}\label{sec:background}

The primary background to the signal comes from prompt SM diphoton production. An additional and reducible source of background occurs when a fragmenting jet mimics a genuine photon signature in the detector. Two different methods are used to determine the SM background. As described in Section~\ref{sec:bkg_resonant}, the resonant signal search uses a maximum likelihood fit to the diphoton invariant mass spectrum, seeking a local excess in the data that could indicate the presence of beyond-SM physics. The nonresonant signal search uses simulation to model the SM diphoton component and a method based on control samples in data to estimate the contribution from misidentified jets; this method is described in Section~\ref{sec:bkg_nonresonant}.

\subsection{Background for resonant diphoton search}\label{sec:bkg_resonant}

In the resonant signal search, the background \mgg spectrum is described by a parametric function of \mgg{}:
\begin{linenomath}
\begin{equation}
	\label{eqn:mgg_formula}
	f(\mgg) = \mgg^{a + b \, \log( \mgg / {\GeVns} )},
\end{equation}
\end{linenomath}
where the parameters $a$ and $b$ are obtained from a fit to the data and are considered as
unconstrained nuisance parameters in the hypothesis test.

The chosen parametric form is designed to be an approximation of the true but unknown background shape. The degree of accuracy with which the model describes the true shape is tested using a set of five different parametric models, all of which can describe relatively well the observed \mgg spectrum. If the chosen model is flexible enough to accommodate the shape of each of the alternative models, then we assume that it would likely be able to describe a similar true background shape.

The accuracy of the background determination is assessed with data using the following procedure: five different parametric background models are fit to the data in order to build five different truth models, and the accuracy of the chosen parametrization is quantified by studying the difference between the true and predicted number of background events in several \mgg intervals in the search region. The width of these intervals ranges between 10 and 500\GeV, in order to keep an equal amount of data in each of the mass ranges. The five functions are chosen from five different families of functions. For each family, the function that requires the minimum number of parameters to fit the data with a $\chi^2$ probability greater then 5\% is chosen. Pseudo-experiments are drawn from the mass spectrum predicted by the different background models. The total number of events in each pseudo-experiment is taken from a Poisson distribution where the mean is determined by the observation in data. For each interval, the distribution of the pull variable, defined as the difference between the true and predicted number of events  divided by the estimated statistical uncertainty, is constructed. If the absolute value of the median of this distribution is found to be above 0.5 in a window, an additional uncertainty is assigned to the background parametrization. A modified pull distribution is then constructed increasing the statistical uncertainty on the fit by an extra term, denoted ``bias term'', which is parametrized as a smooth function of \mgg, tuned so that the absolute value of the median of the modified pull distribution is below 0.5 for all regions. The additional uncertainty is then included in the likelihood function by adding to the background model a component having the same shape as the signal, with a normalization coefficient distributed as a Gaussian function of mean zero and width equal to the integral of the bias term over the FWHM of the tested signal shape. The inclusion of the additional component has the effect of avoiding falsely positive or negative tests that could be induced by a mismodeling of the background shape, and it degrades the analysis sensitivity by 0.1 to 10\% depending on the mass and width of the signal hypothesis.

\subsection{Background for nonresonant diphoton search}\label{sec:bkg_nonresonant}

In the search for nonresonant deviations from the SM diphoton spectrum, we make a prediction of the invariant mass distribution expected from SM diphoton events, as well as of contributions from photon+jet or dijet events where one or two jets fragment in such a way as to resemble a photon signature in the CMS detector. Prompt SM diphoton production can occur via quark annihilation or gluon fusion processes. This irreducible background is modeled using the MC event generator \SHERPA~2.1.1~\cite{Gleisberg:2008ta,Hoeche:2009xc}. Up to three additional jets are added to the diphoton final state, to better simulate phase space regions with small-to-medium angular separation between the two photons. The CT10 set of PDFs is used for generation, and the generated events are subject to a simulation of the CMS detector response based on the \GEANTfour package~\cite{Agostinelli:2002hh}. While the inclusion of explicit final-state jets in the \SHERPA sample incorporates the real radiation component of higher-order corrections to the basic diphoton process, virtual corrections are not included in these \SHERPA simulated events. This is remedied by performing a full NNLO calculation of SM diphoton production using \MCFM~8.0~\cite{Campbell:2016yrh}. A $K$ factor, defined as the ratio of the \MCFM NNLO prediction to the \SHERPA prediction, is calculated as a function of the diphoton invariant mass. This $K$ factor is then used to reweight and correct the \SHERPA events (after detector simulation) to obtain the prediction for the SM diphoton invariant mass spectrum from two real photons. The $K$ factor is calculated separately for the EBEB and EBEE acceptance categories, with the renormalization, factorization, and fragmentation scales for the calculation set to be \mgg. Events with very small $\DR_{\gamma\gamma}$ ($<$0.45) are rejected to avoid an infrared divergence. The $K$ factor for the EBEB (EBEE) event category varies from 1.4--1.8 (1.5--2.0) over the range 0.5--2.0\TeV. Next-to-leading-order (NLO) $K$ factors are also computed in a similar way with both \MCFM and \textsc{diphox}~\cite{Binoth:1999qq}, and are used to extract systematic uncertainties in the shape of the central, NNLO calculation.

An additional background occurs from SM photon+jet or dijet events when one or two jets fragment in such a way as to resemble a photon signature in the detector. Application of shower shape and isolation criteria to the identification of photon candidates as described in Section~{\ref{sec:event_selection}} aims to minimize this reducible background, and we use a data-driven method to estimate the remaining contribution. The method assumes that events where a jet has fragmented to produce a fake photon signature in the detector can be modeled by other events where the jet fragmented differently. We define two distinct categories of jet fragmentation objects, and measure the ratio of the yields of objects in these two fragmentation categories, in a jet-triggered reference data sample.

In the first category, whose objects are counted for the numerator of the ratio, are those misidentified jets that pass the photon identification requirements; these are the ones whose contribution to the background we wish to determine. When measured in a reference data sample, however, genuine photons passing the same ID requirements contribute also to this category. This contribution is subtracted out statistically, in the following way. The shower shape variable \sieie is used as a discriminant between genuine photons and jets misidentified as photons. A template is constructed for the \sieie distribution of real photons using simulated events, and a \sieie template for jets misidentified as photons is obtained from a control sample in data enriched in such jets, by inverting the identification requirement for charged hadron isolation. These templates are then fit to the observed \sieie distribution of the numerator objects, to determine the relative contribution of genuine photons, which is then subtracted. This template-fitting and subtraction procedure is performed in bins of photon \pt since the \sieie templates are found to be \pt dependent. After this procedure, what remains is the contribution to the numerator object category only from jets that have been misidentified as photons.

The denominator category consists of `photon-like' jets that pass a less strict version of the photon ID, but still have a high electromagnetic energy component. They are additionally required to fail at least one of the isolation or shower shape requirements for photon candidate identification, ensuring that the two fragmentation categories are mutually exclusive and that there is negligible contamination from real photons in the denominator category.

The ratio of the numbers of objects in the two jet fragmentation categories is measured as a function of the photon \pt in an independent jet-triggered data sample. The background prediction for jets that are misidentified as photons is then obtained by considering events containing objects that pass the looser `photon-like' jet definition, and reweighting those events by the relative fragmentation ratio. This contribution from misidentified jets to the diphoton spectrum is found to decrease within the mass range $0.5 < \mgg <1\TeV$ from 9 to 4\% of the total background in the EBEB category and from 28 to 17\% in the EBEE. For $\mgg > 1\TeV$, this integrated contribution from jets misidentified as photons is predicted to be less than 4 (14)\% in the EBEB (EBEE).

\section{Systematic uncertainties}\label{sec:systematics}

\subsection{Systematic uncertainties for the resonant search}\label{sec:systematics_ree}

The systematic uncertainties on the search for resonant diphoton signals are smaller than the associated statistical uncertainties. The parametric background model has no associated systematic uncertainties, except for the bias term uncertainty described previously. The shape coefficients are treated as unconstrained nuisance parameters, and thus the associated uncertainties contribute to the statistical uncertainty.

Uncertainties associated with the signal modeling are:
\begin{itemize}
	\item a 2.5\% uncertainty in the signal normalization assigned to reflect the uncertainty in the total integrated luminosity~\cite{CMS-PAS-LUM-17-001};
	\item a 6\% uncertainty in the signal normalization to reflect uncertainty in the selection efficiency;
	\item a 6\% uncertainty in the signal normalization in order to account for the variation in the kinematic acceptance estimated by comparing the use of the alternative PDF sets CT10~\cite{Lai:2010vv,Gao:2013xoa}, NNPDF2.3~\cite{Ball:2012cx}, and MSTW08~\cite{Martin:2009iq} on the signal hypothesis and taking the largest deviation, following the PDF4LHC recommendations~\cite{Alekhin:2011sk,Botje:2011sn};
	\item a 1\% uncertainty in the photon energy scale is included in the fit to take into account the uncertainty associated with the photon energy scale at the \PZ boson mass and its extrapolation to higher masses; and
	\item the uncertainty in the photon energy resolution correction factor evaluated by summing and subtracting 0.5\% in quadrature from the estimated additional Gaussian smearing measured at the \PZ boson peak.
\end{itemize}

\subsection{Systematic uncertainties for the nonresonant search}\label{sec:systematics_nonres}

Although the SM diphoton background prediction is at NNLO accuracy, there is still the possibility of contribution from unaccounted-for higher-order terms. Therefore, we allow the normalization of the predicted diphoton background to float freely, constrained only by the data (predominantly at low \mgg where the statistical uncertainty is smallest).  Floating the normalization also absorbs sources of uncertainty associated with the integrated luminosity measurement, trigger, and photon selection efficiency. Provided the shape of the signal differs significantly from the background, the analysis will still discriminate between the two.  Scale uncertainties in the $K$ factor calculation are estimated by simultaneously varying the renormalization, factorization, and fragmentation scales between $\mgg/2$ and $2\mgg$.  The difference between the shape of the NNLO \MCFM \mgg spectra and the shape predicted at NLO by \MCFM and \textsc{diphox} is also taken as an uncertainty and as a conservative bound on the effects of the absent higher-order terms.

Systematic uncertainties associated with the PDFs are calculated using \textsc{diphox} at NLO with a consistent set of NLO CT10 PDFs.  The 26 eigenvectors of this PDF set are varied individually by $\pm1$ standard deviation (rather than taking an envelope), which allows us to treat consistently the correlations of the uncertainties as a function of \mgg and between the EBEB and EBEE categories. Uncertainties due to the photon energy scale and resolution have a negligible impact on the nonresonant search. The uncertainty in the misidentification rate for both the barrel and endcap is approximately 30\%, as estimated from the variation of the fake rate as a function of pileup and photon $\eta$, and from the degree of variation observed in a test of the method using MC simulation. A separate shape uncertainty due to differences in the samples from which the fake rate was constructed (using a multijet- or dimuon-triggered data set) makes a subdominant contribution.

Uncertainties in the normalization of the extra dimensional signal from the measured integrated luminosity and photon selection efficiency are 2.5 and 6\%, respectively, in agreement with the resonance search.  Systematic uncertainties in the signal shape from PDFs are treated in the same manner as the background by separately varying the 26 eigenvectors. These uncertainties are assumed to be 100\% correlated between signal and background.  The effect of the PDF uncertainty on the signal cross section is treated as a theoretical uncertainty and is not propagated in the upper limits.

\section{Results}\label{sec:results}

\subsection{Results of the search for resonant excesses}\label{sec:results_res}

\begin{figure*}[!htb]
    \centering
    \includegraphics[width=0.48\textwidth]{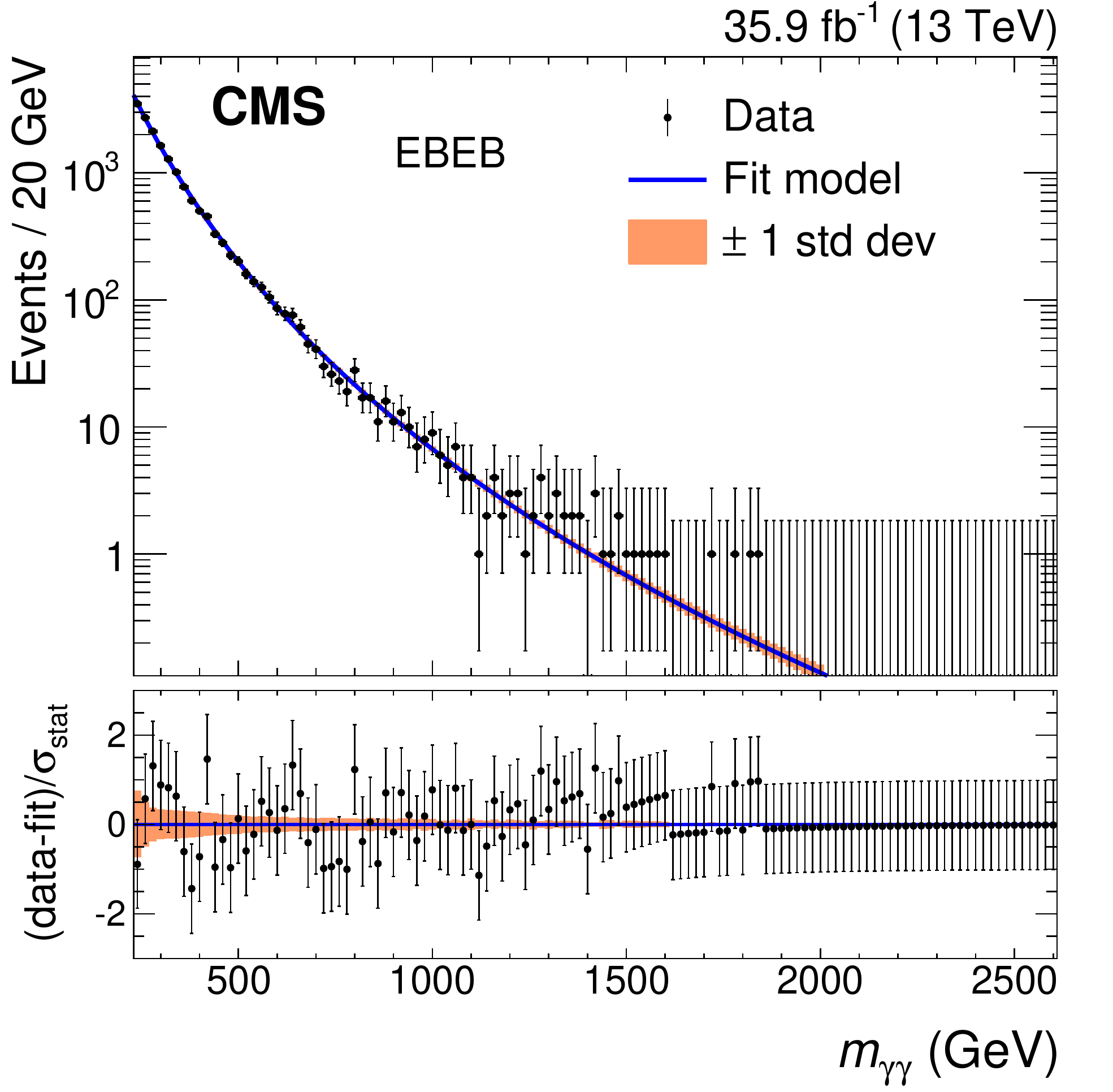}
    \includegraphics[width=0.48\textwidth]{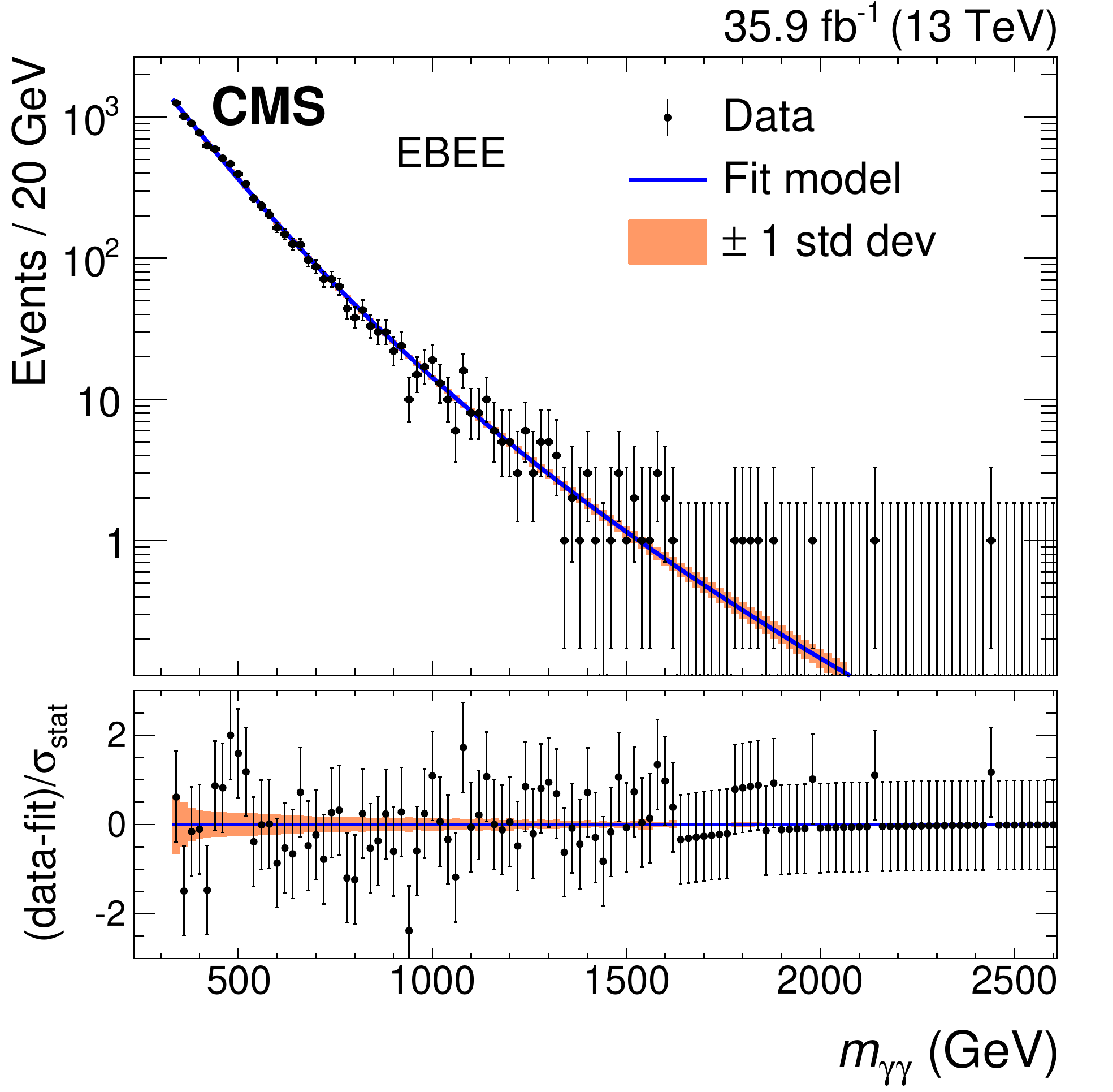}
    \caption{
      Observed diphoton invariant mass spectra for the EBEB (left) and EBEE (right) categories. Also shown are the results of a likelihood fit to the background-only hypothesis. The shaded region shows the one standard deviation uncertainty band associated with the fit, reflecting the statistical uncertainty of the data. The lower panels show the difference between the data and fit, divided by the statistical uncertainty in the data points.
    }
    \label{fig:fits}
\end{figure*}

The \mgg distribution of the selected diphoton events and the background parametrization obtained through an unbinned maximum likelihood fit to these events are shown in Fig.~\ref{fig:fits}. This parametric form corresponds to the one chosen to model the background given by Eq.~(\ref{eqn:mgg_formula}), as detailed in Section~\ref{sec:background}.

The results of the search are interpreted in the framework of a composite statistical hypothesis test. A simultaneous fit to the invariant mass spectra of the EBEB and EBEE event categories is used to study the compatibility of the data with the background-only and the signal+background hypotheses.

The test statistic used in the hypothesis tests are based on the profile likelihood ratio:
\begin{linenomath}
\begin{equation}\label{eqn:mgg_likelihood}
	q(\mu) = -2 \log \frac{ L(\mu S + B | \hat\theta_{\mu} ) } {L (\hat\mu S + B | \hat\theta )},
\end{equation}
\end{linenomath}
where $S$ and $B$ are the probability density functions for the resonant diphoton signal production process and the SM background, respectively; $\mu$ is the signal strength parameter, defined as the ratio between the measured and expected signal cross sections; and $\theta$ are the nuisance parameters of the model used to account for the associated systematic uncertainties. The notation $\hat{x}$ indicates the best fit value of the parameter $x$, while $\hat{x}_y$ denotes the best fit value of $x$, conditionally on $y$.

The data are in agreement with the absence of any significant resonant excess of events. The largest deviation observed is an approximately 2 standard deviation local excess at 1.2\TeV for the wide-width hypothesis, and is similar for both the spin-0 and spin-2 signals.

To set upper limits on the resonant diphoton production rate, the modified frequentist method, commonly known as \CLs~\cite{CLS1,CLS2}, is used following the prescription described in Ref.~\cite{CMS-NOTE-2011-005}. Asymptotic formulas~\cite{Cowan:2010js} are used in the calculations of limits.

\begin{figure*}[!htb]
    \centering
    \includegraphics[width=0.48\textwidth]{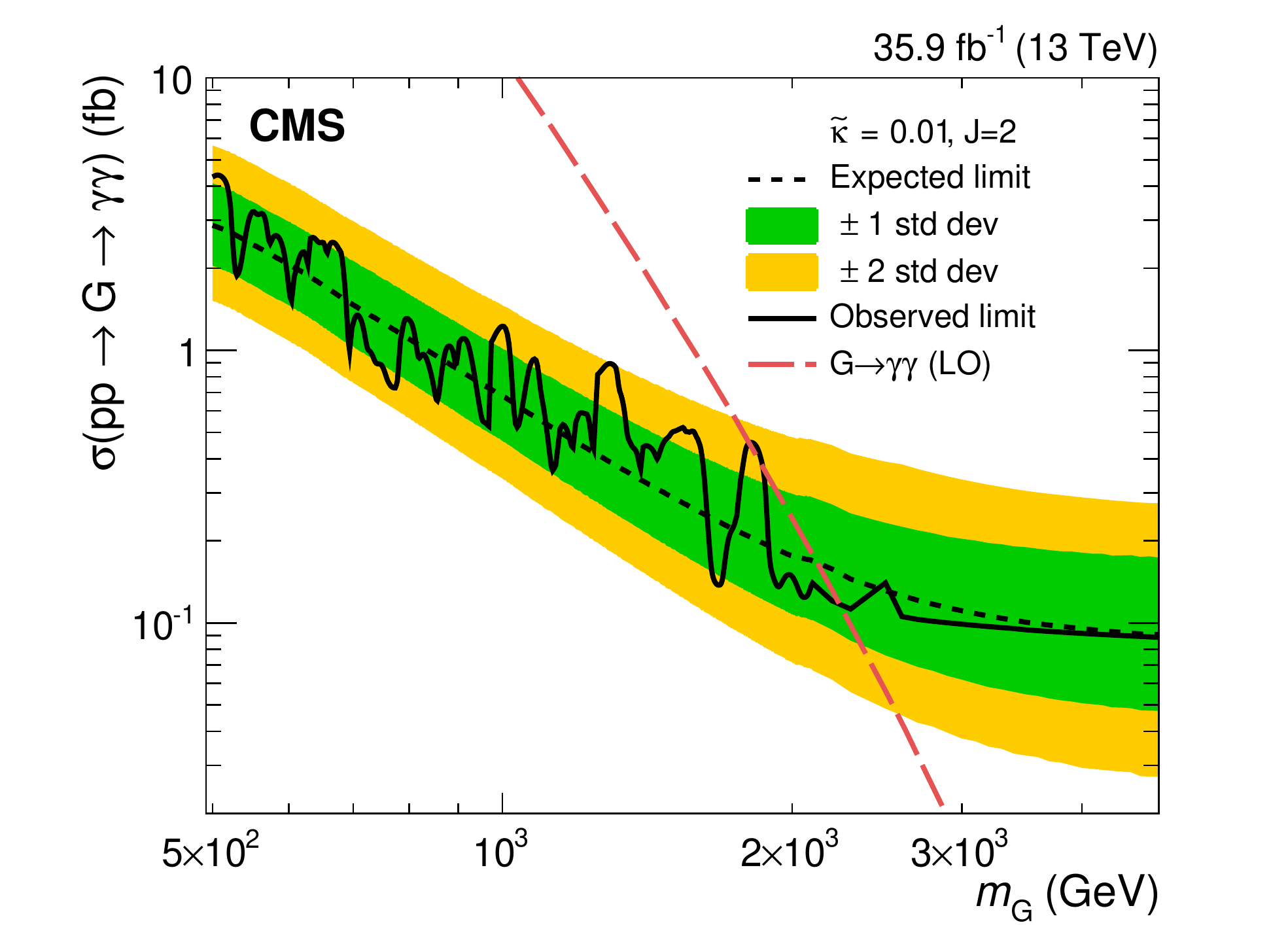}
    \includegraphics[width=0.48\textwidth]{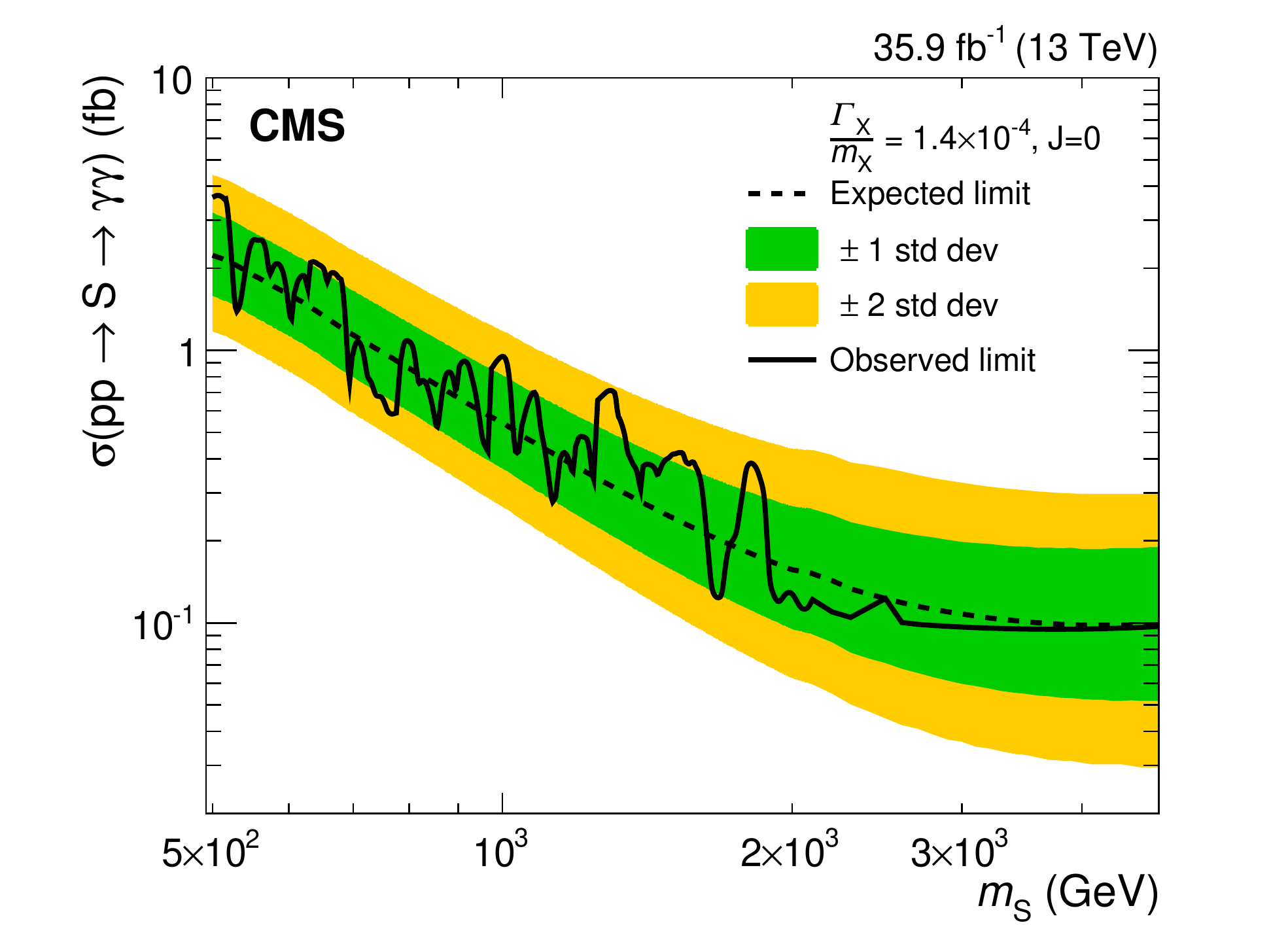} \\
    \includegraphics[width=0.48\textwidth]{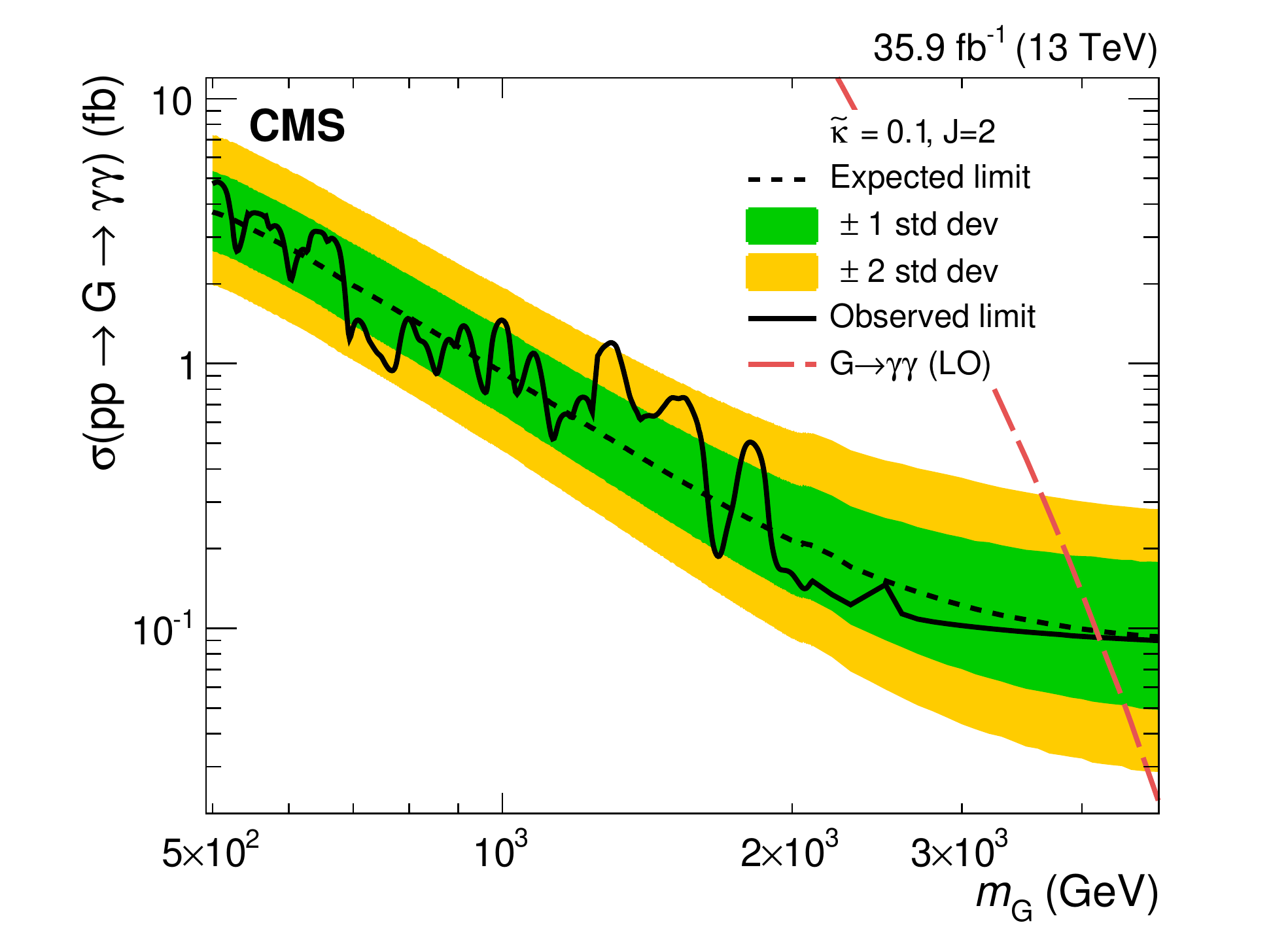}
    \includegraphics[width=0.48\textwidth]{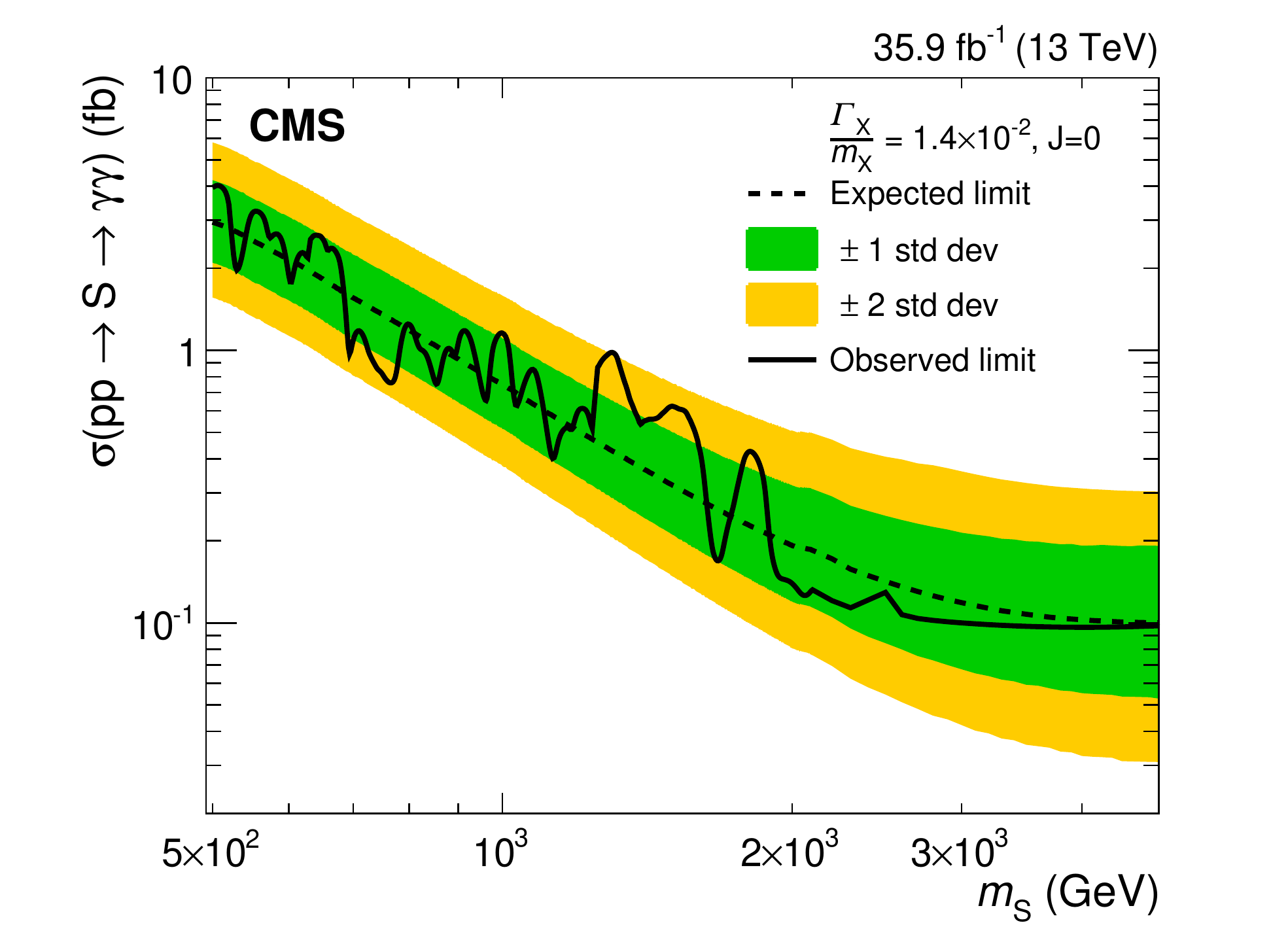} \\
    \includegraphics[width=0.48\textwidth]{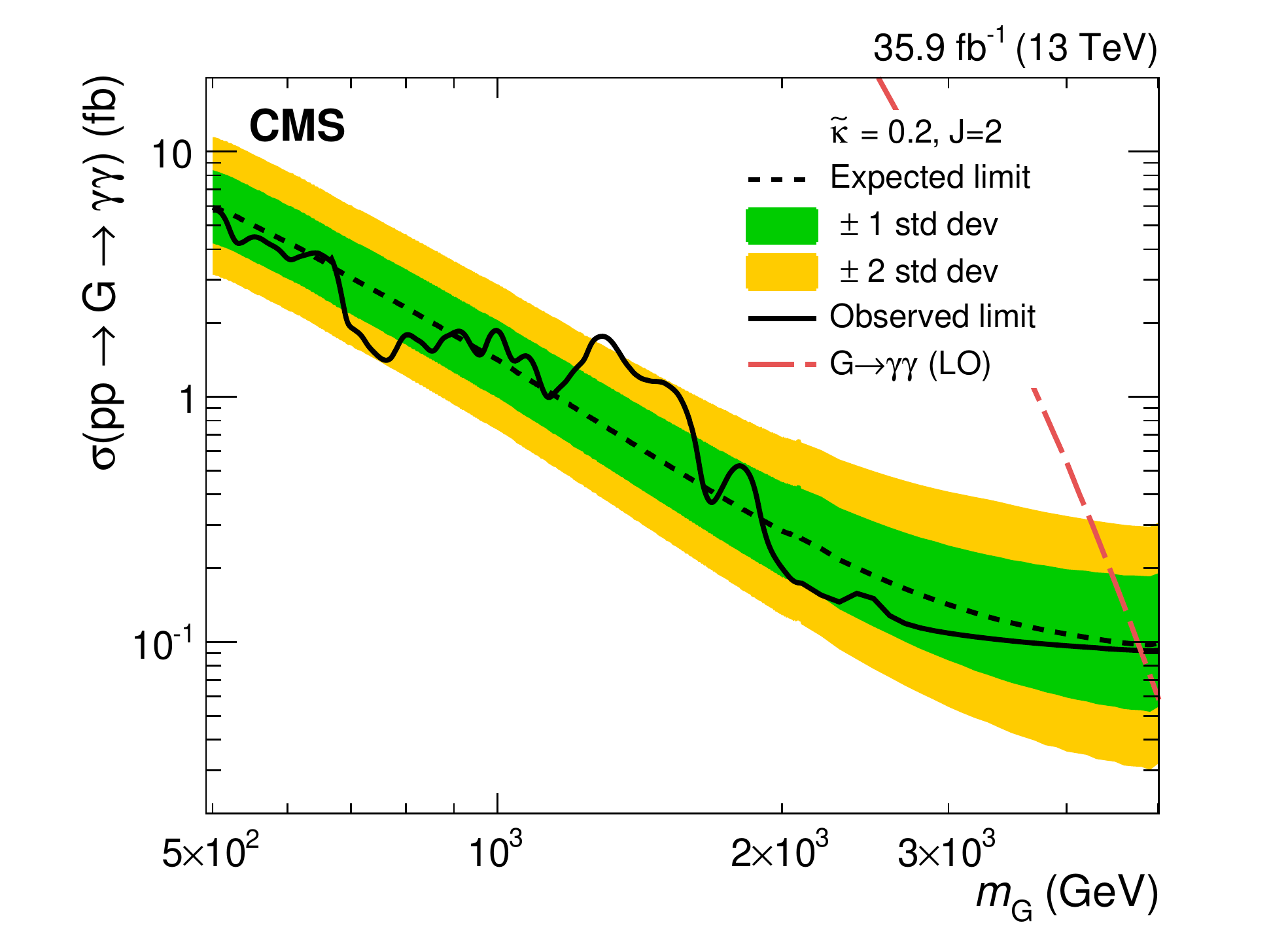}
    \includegraphics[width=0.48\textwidth]{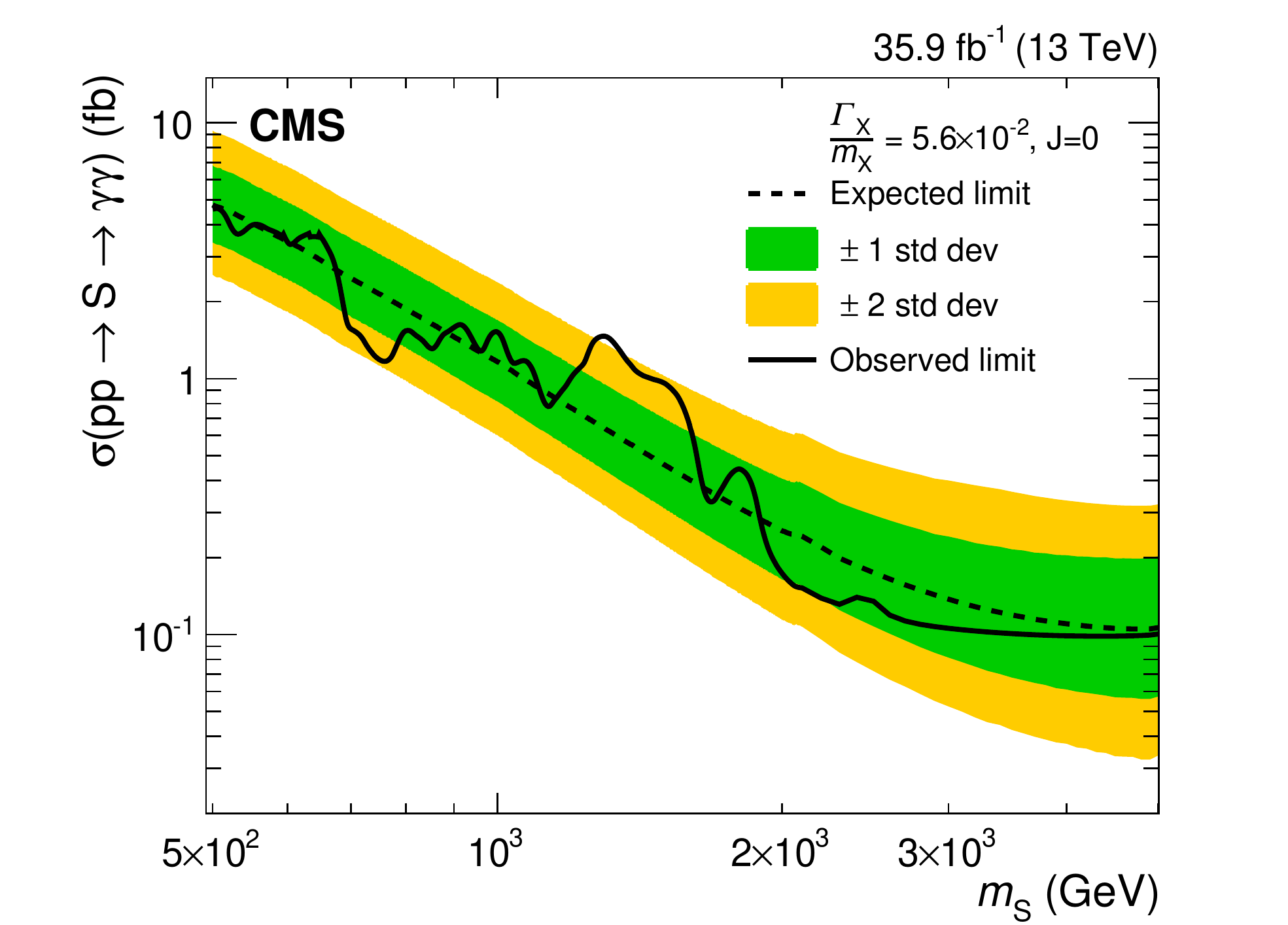}
    \caption{
      Expected and observed 95\% \CL upper limits on the production cross section for RS gravitons of mass \mG and three values of \ktild (left) and for spin-0 resonances of mass \mS produced via gluon fusion for the three width hypotheses (right). The shaded bands represent the 1 and 2 standard deviation uncertainty in the expected limit.
    }
    \label{fig:limits_spin0_spin2}
\end{figure*}

Expected and observed upper limits at 95\% confidence level (\CL) on the production of scalar and RS graviton resonances are shown in Fig.~\ref{fig:limits_spin0_spin2}. Using leading order cross sections from \PYTHIA, RS gravitons with masses \mX below 2.3, 4.1, and 4.6\TeV can be excluded for $\ktild=0.01$, 0.1, and 0.2, respectively, corresponding to $\GammaXOmX=1.4\times10^{-4}$, $1.4\times10^{-2}$, and $5.6\times10^{-2}$, respectively.

Limits can also be set, in a model independent fashion, on the cross sections for events in the fiducial volume for the resonant $\Pp \Pp \to \gamma\gamma$ process. The signal shape is modeled in the same way as for the benchmark models, while the signal normalization accounts only for the detector efficiency and not for any particular signal acceptance. The fiducial volume is defined by selecting events in which both photons have generator-level $\pt > 75\GeV$, to match the selection applied in the event reconstruction and selection. Generator-level photons are also required to have an isolation energy of less than 10\GeV in a cone of radius 0.4 around the photon direction. The isolation energy is defined as the \pt sum of all final state particles except neutrinos and the signal photon itself.

\begin{figure*}[!htb]
    \centering
    \includegraphics[width=0.48\textwidth]{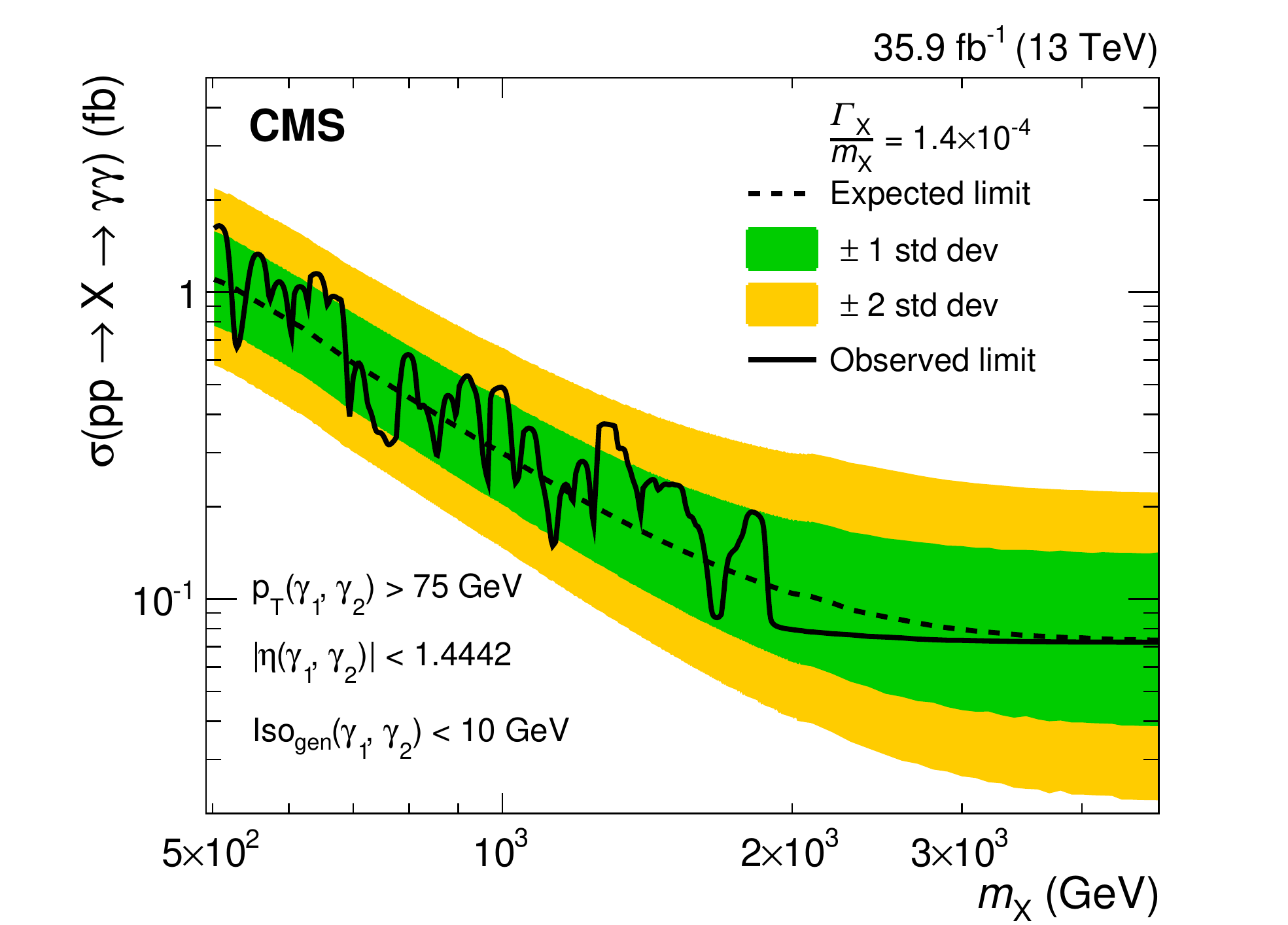}
    \includegraphics[width=0.48\textwidth]{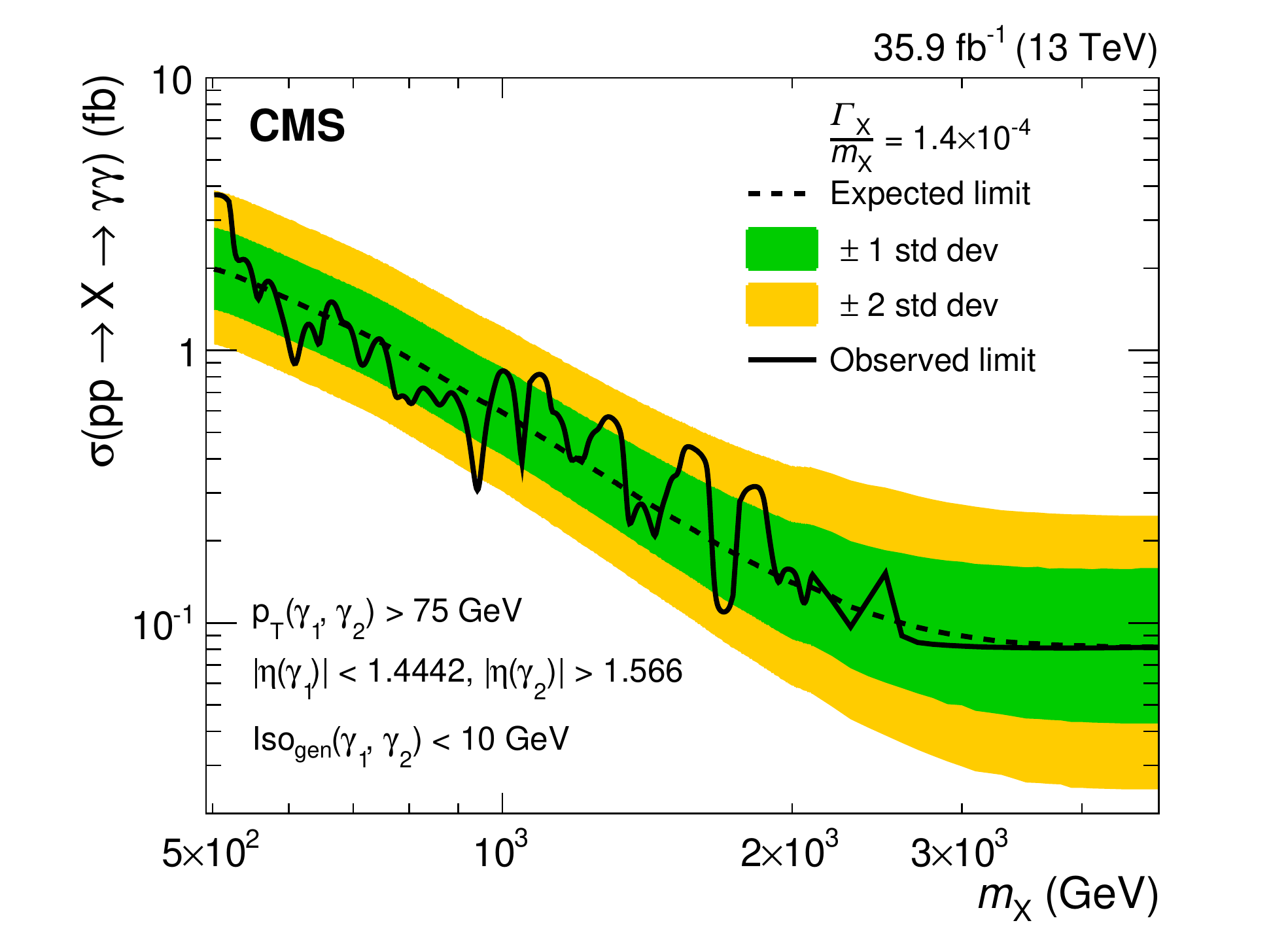} \\
    \includegraphics[width=0.48\textwidth]{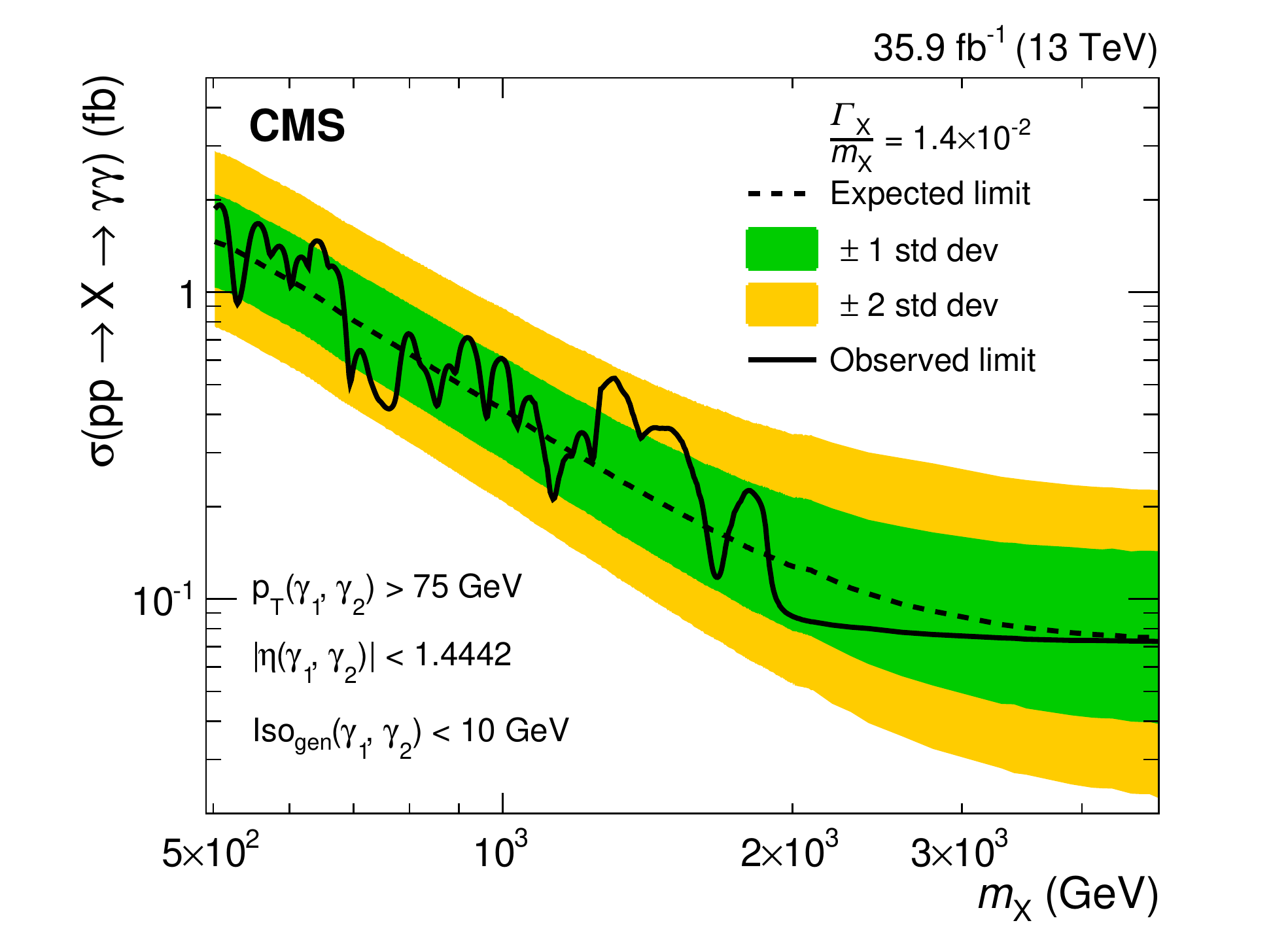}
    \includegraphics[width=0.48\textwidth]{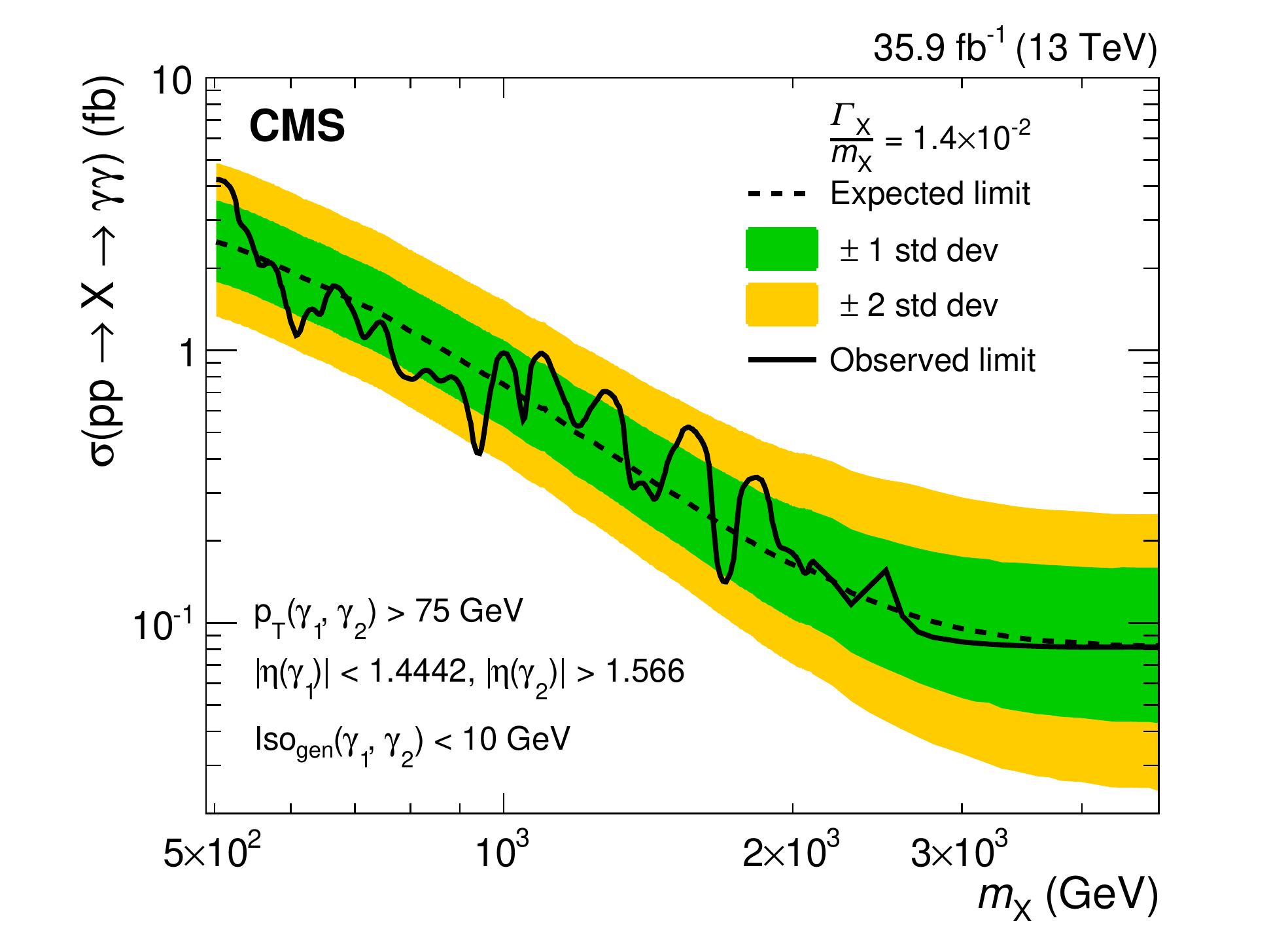} \\
    \includegraphics[width=0.48\textwidth]{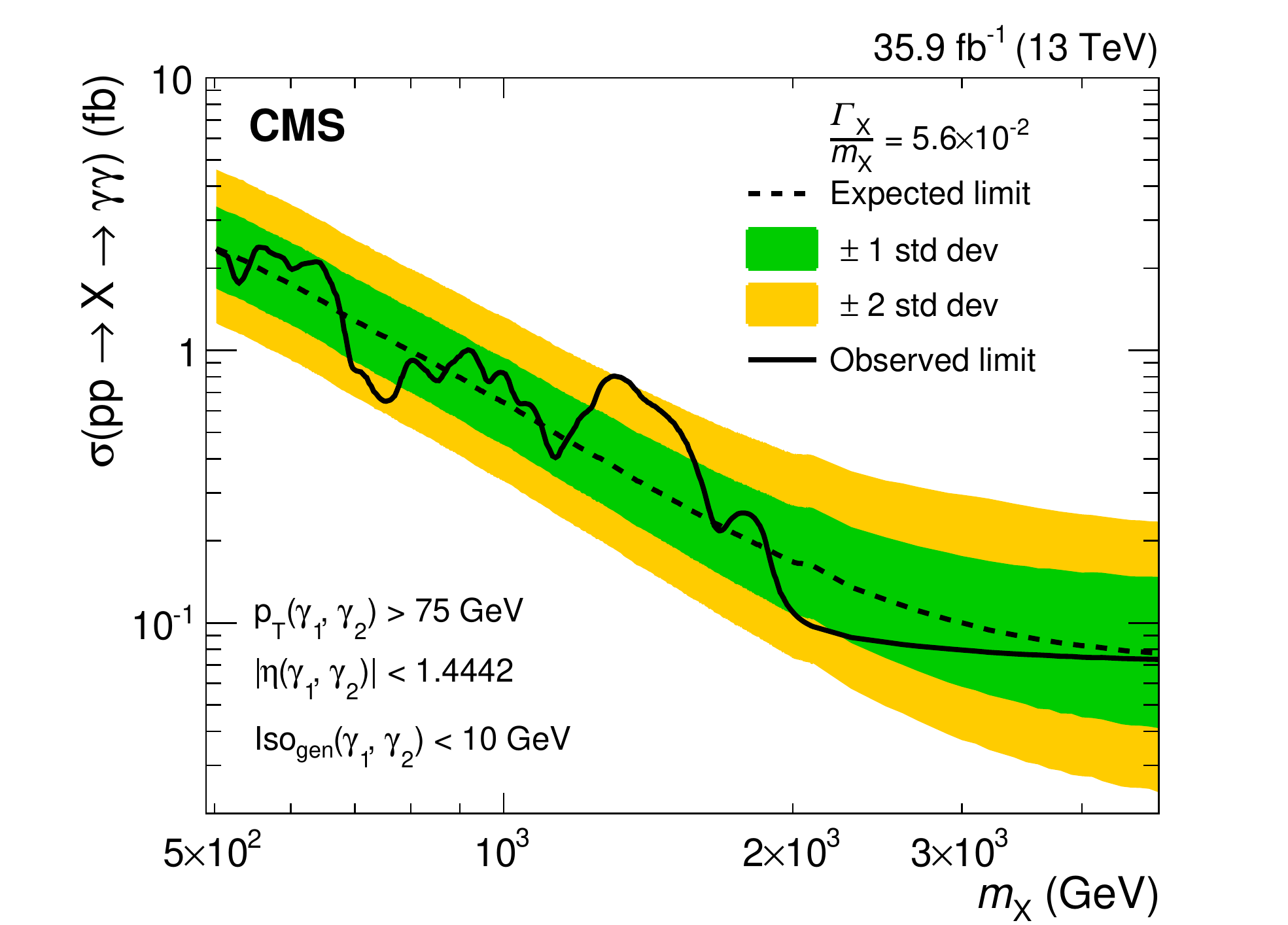}
    \includegraphics[width=0.48\textwidth]{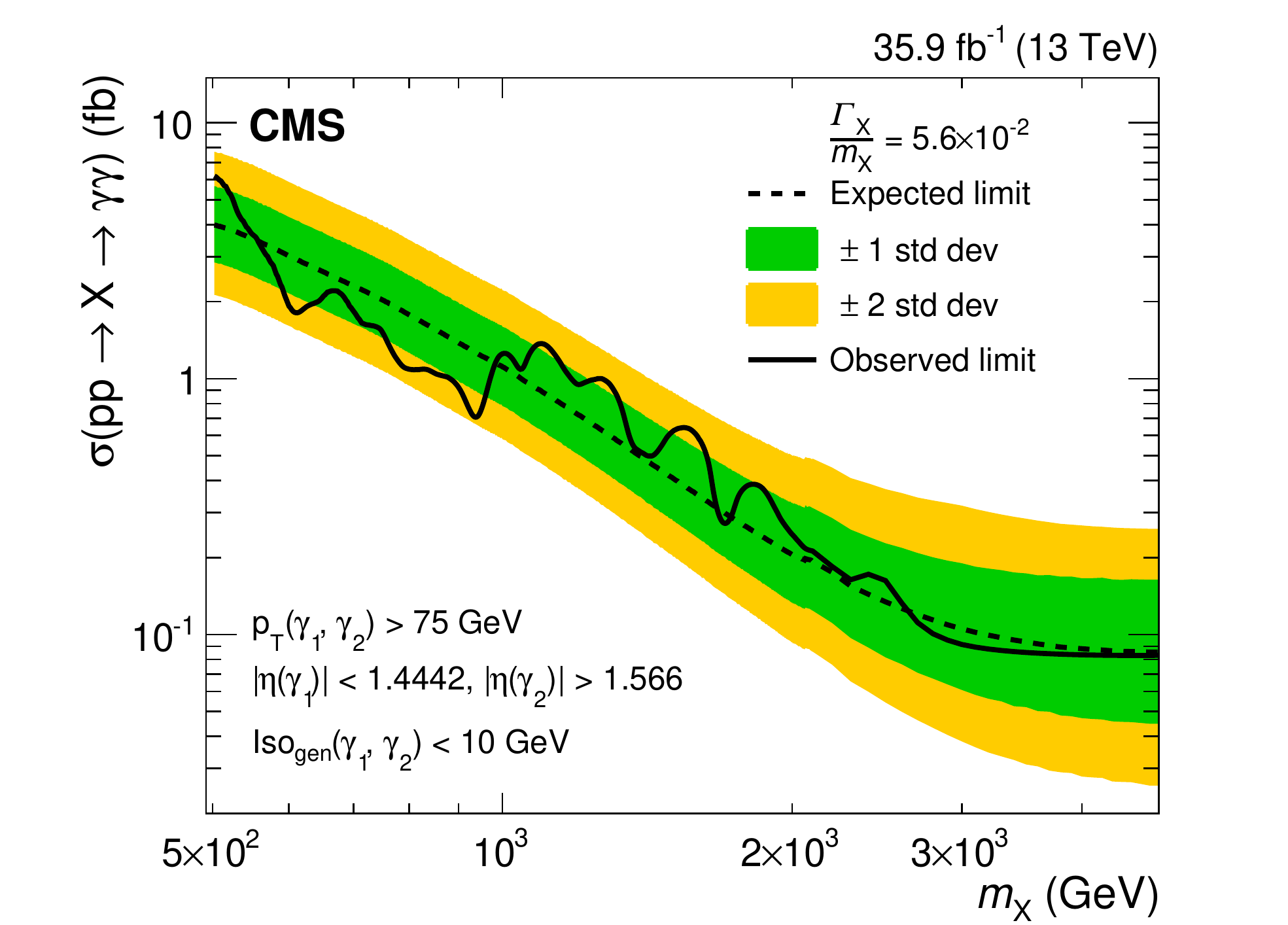}
    \caption{
      Expected and observed 95\% \CL upper limits on the fiducial cross section for the resonant $\Pp\Pp \to \gamma\gamma$ process. Shown are the results in the EBEB (left) and EBEE (right) categories for the three width hypotheses. The shaded bands represent the 1 and 2 standard deviation uncertainty in the expected limit.
    }
    \label{fig:fiducial_limits}
\end{figure*}

The fit is performed independently in the EBEB and EBEE categories for each of the following width hypotheses: $\GammaXOmX=1.4\times10^{-4}$, $1.4\times10^{-2}$, and $5.6\times10^{-2}$. The results for the median expected exclusion limits on the fiducial cross sections are presented in Fig.~\ref{fig:fiducial_limits} for each width hypothesis.

\subsection{Results of the search for nonresonant excesses}\label{sec:results_nonres}

In the nonresonant search, constraints on the signals are determined by adopting a Bayesian statistical approach. A binned likelihood is constructed from the data, with the binning in 100\GeV steps beginning at 500\GeV. The signal strengths for each model are assumed to have a flat prior (bounded below by zero).  Nuisance parameters are assigned to each of the systematic uncertainties described in Section~\ref{sec:systematics_nonres}.  Except for the diphoton background normalization nuisance parameter, which has a flat prior, the prior shapes are either lognormal or Gaussian, depending on whether or not the uncertainty is bounded below by zero.  Nuisance parameters are marginalized using a Markov chain MC method~\cite{Metropolis:1953am}.

\begin{figure*}[!htb]
	\centering
 	\includegraphics[width=0.95\textwidth]{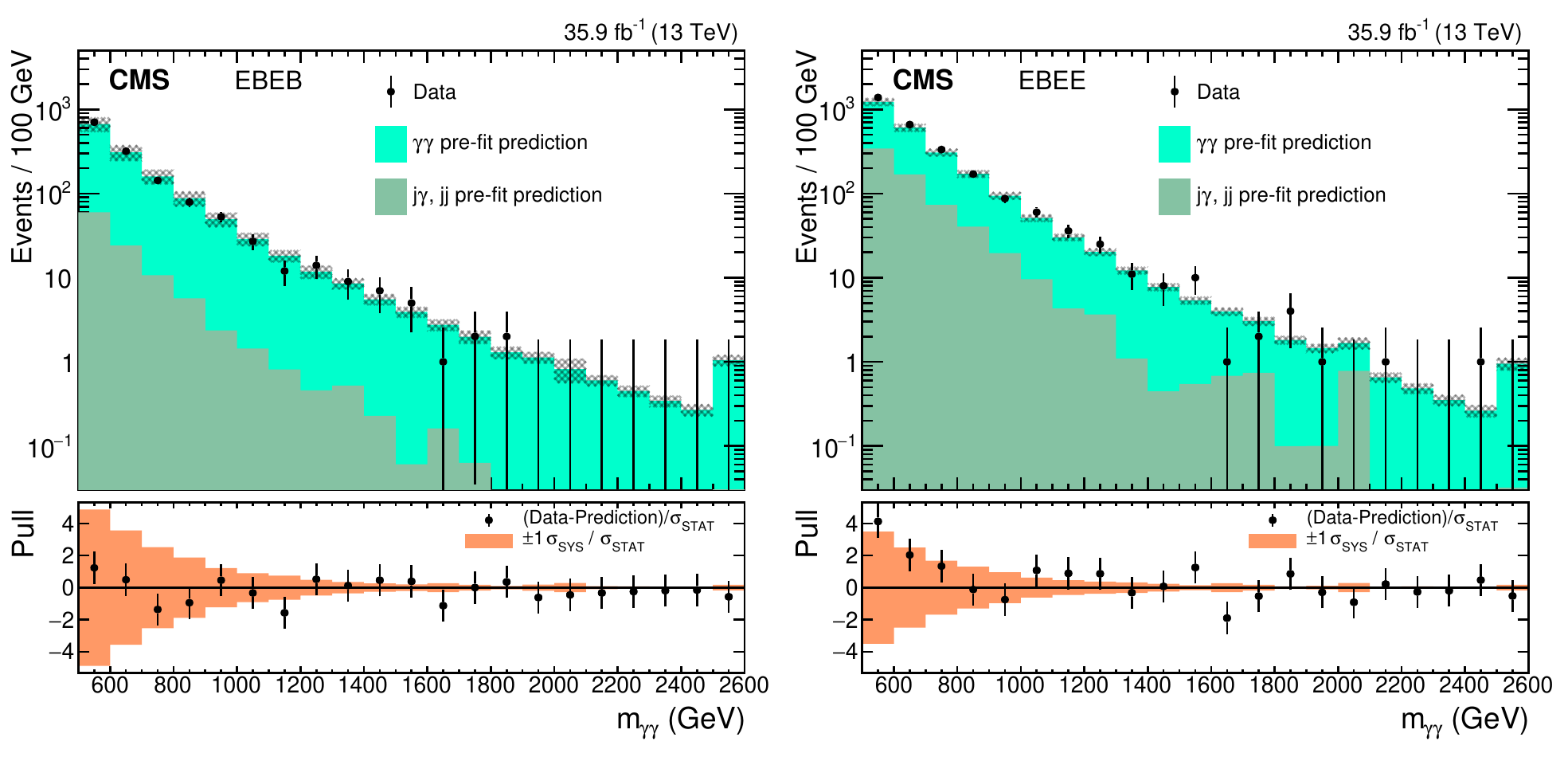}\\
  	\includegraphics[width=0.95\textwidth]{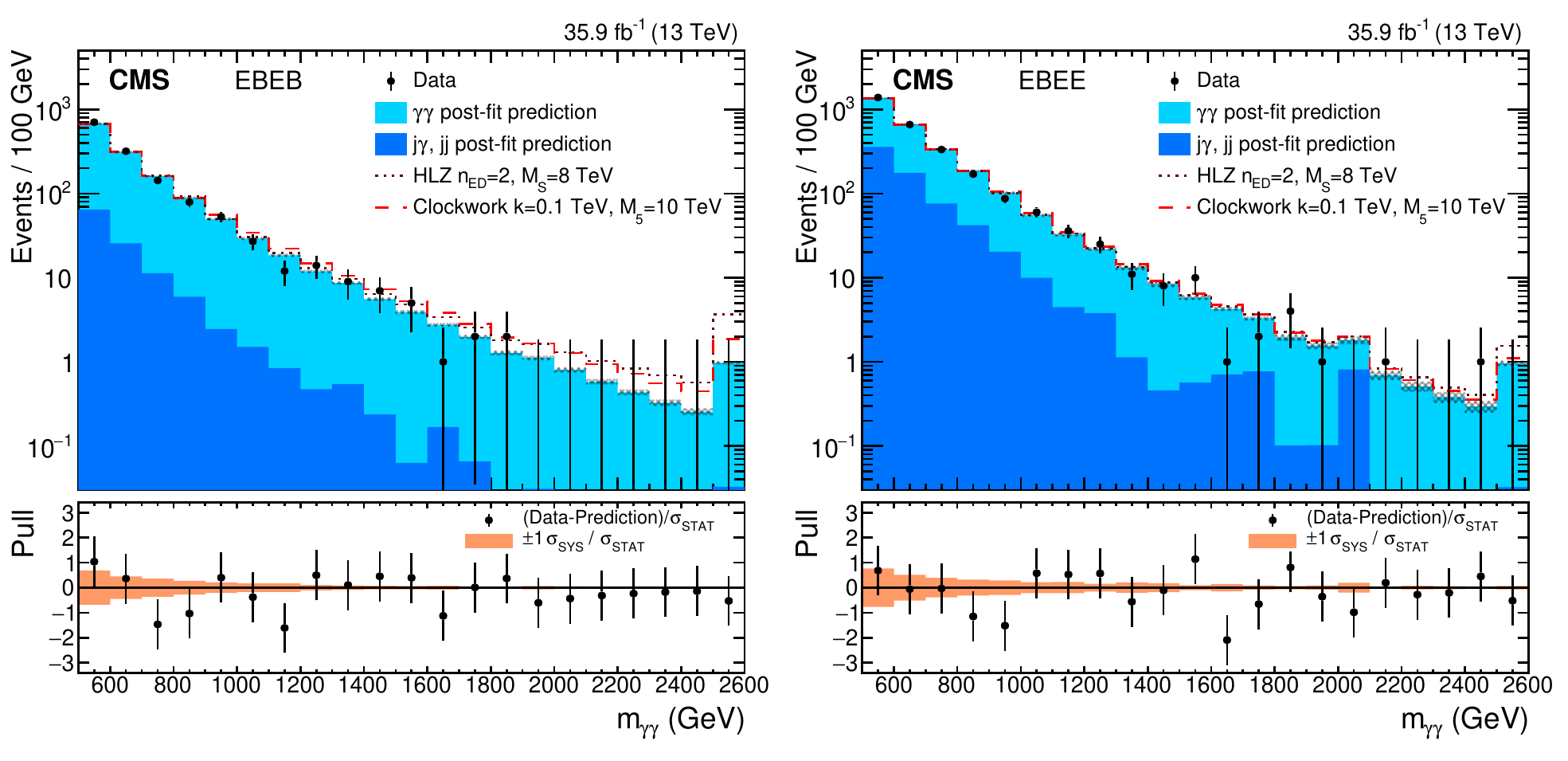}
  	\caption{
  		The diphoton invariant mass distributions in the EBEB (left) and EBEE (right) categories for the SM diphoton background prediction and the fake background measurement compared to the data. The last bin includes the overflow. The error bars on the points indicate the statistical uncertainty. The upper (lower) plots show the pre-fit (post-fit) background estimates. The hatched bands indicate the total pre- or post-fit systematic uncertainties. Invariant mass distributions from two signal scenarios are superimposed on the lower plots. The bottom panels show the pull distributions, indicating the difference between the data and background prediction, divided by the uncertainty in the background, with error bars representing the statistical uncertainty and shaded bands showing the one standard deviation systematic uncertainty, normalized by the statistical uncertainty.
  	}
	\label{Fig:CR_PrePostFits}
\end{figure*}

Figure~\ref{Fig:CR_PrePostFits} (upper) presents the data and background prediction ``pre-fit,'' \ie, before the marginalization of the nuisance parameters.  The shaded bands show the systematic uncertainties, neglecting the (unbounded) normalization of the diphoton prediction and the NLO shapes.  Figure~\ref{Fig:CR_PrePostFits} (lower) presents the data and background prediction ``post-fit,'' \ie, after the marginalization of the nuisance parameters.  Good agreement is found within the uncertainties for both the pre- and post-fit spectra, although the uncertainties are smaller, as expected, in the latter case.

For the ADD model, upper limits are set at 95\% \CL on the signal strength, which are translated into lower limits on the mass scale \MS by interpolating the value of \MS that has a signal strength of 1.0 excluded. Table \ref{Tab:Limits2016} summarizes the results for all ADD model conventions probed. The excluded values of \MS range from 5.6 to 9.7\TeV, depending on the convention.

\begin{table*}[htb]
	\centering
	\topcaption{
		Exclusion lower limits obtained on the mass scale \MS (in units of {\TeVns}) for various conventions used in the calculation of the ADD large extra dimensional scenario, as described in Section~\ref{sec:signals}. The total asymmetric uncertainties are shown on the expected limits.
	}
	\cmsTable{
	\begin{scotch}{cccccccccc}
		\vspace*{-3.5mm} &&&&&&&&& \\
		\multirow{2}{*}{Signal} & GRW & \multicolumn{2}{c}{Hewett} & \multicolumn{5}{c}{HLZ} \\
		& & negative & positive & $\nED=2$ & $\nED=3$ & $\nED=4$ & $\nED=5$ & $\nED=6$ & $\nED=7$ \\ [\cmsTabSkip]
		\hline
		\vspace*{-2.5mm} &&&&&&&&& \\
		Expected & $7.1^{+0.7}_{-0.5}$ & $5.5^{+0.1}_{-0.3}$ & $6.3^{+0.6}_{-0.4}$ & $8.4^{+1.3}_{-1.1}$ & $8.4^{+0.8}_{-0.6}$ & $7.1^{+0.7}_{-0.5}$ & $6.4^{+0.6}_{-0.5}$ & $6.0^{+0.6}_{-0.4}$ & $5.6^{+0.6}_{-0.4}$ \\ [\cmsTabSkip]
		Observed & 7.8 & 5.6 & 7.0 & 9.7 & 9.3 & 7.8 & 7.0 & 6.6 & 6.2 \vspace*{1.0mm} \\
	\end{scotch}
	}
	\label{Tab:Limits2016}
\end{table*}

\begin{figure}[!htb]
	\centering
	\includegraphics[width=\cmsFigWidth]{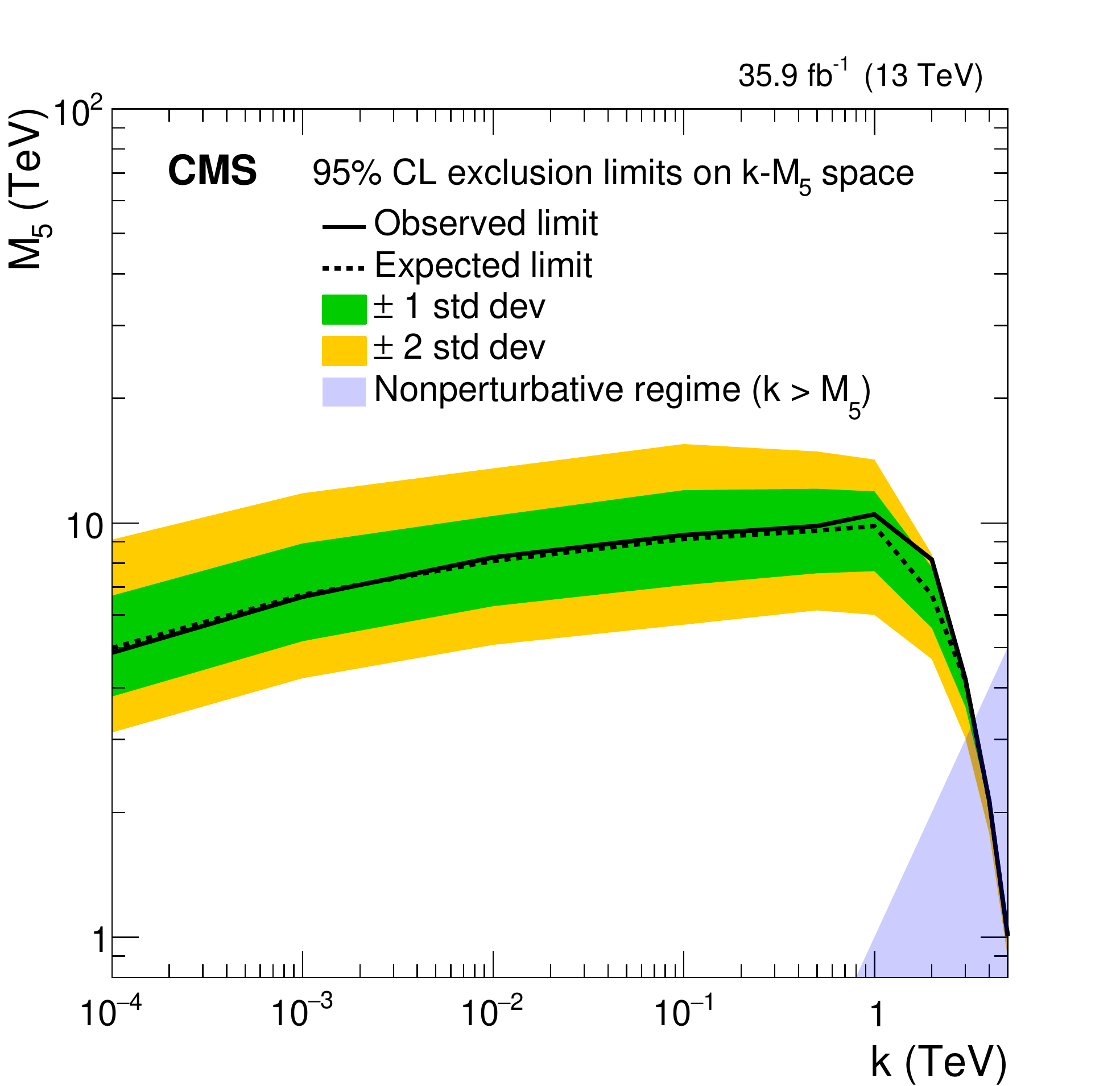}
	\caption{
		The 95\% \CL exclusion limits for the continuous graviton model in the clockwork framework over the $k$--$M_5$ parameter space. The shaded bands represent the 1 and 2 standard deviation uncertainty in the expected limit. The shaded region with $k > M_5$ denotes the area where the theory becomes nonperturbative.
	}
	\label{Fig:LIMIT_Clockwork}
\end{figure}

The limit-setting strategy for the clockwork model is similar to that for the ADD model. Only the portion of the diphoton invariant mass spectrum with $\mgg>900\GeV$ can be used for limit setting on the clockwork model. This constraint is imposed to maintain a statistically precise prediction after the translation of the ADD to clockwork signal. The ADD signal is only simulated for $\mgg>500\GeV$ and the constructed clockwork signal acquires sufficient statistics only above 900\GeV. A simplifying effect is that the signal strength normalization scales as $M_5^{-3}$, so a direct translation between an upper limit on the signal strength and a lower limit on $M_5$ can be made (for a fixed value of $k$). Figure~\ref{Fig:LIMIT_Clockwork} shows the 95\% \CL exclusion limits in the $k$--$M_5$ plane. We are able to exclude values of $M_5$ lower than 5\TeV for $k$ values in the range of 0.2\GeV to 2.0\TeV. The parameter space with $k > M_5$ is excluded by a perturbativity requirement, and this region is denoted `nonperturbative' in the exclusion plot.

\section{Summary}\label{sec:concl}

A search has been performed for physics beyond the standard model in high-mass diphoton events from proton-proton collisions at a center-of-mass energy of 13\TeV. The data used correspond to an integrated luminosity of 35.9\fbinv collected by the CMS detector in 2016. A resonant peak in the diphoton invariant mass spectrum could indicate the existence of a new scalar particle, such as a heavy Higgs boson, or of a Kaluza--Klein excitation of the graviton in the Randall--Sundrum model of warped extra dimensions. A nonresonant excess could be a signature of large extra dimensions, in the scenario by Arkani-Hamed, Dimopoulos, and Dvali, or a continuum clockwork model.

The data are found to be in agreement with the predicted background from standard model sources, and no evidence for new physics is seen. Masses below 2.3--4.6\TeV are excluded at 95\% confidence level for the excited state of the Randall--Sundrum graviton, for a coupling parameter in the range $0.01 < \tilde{k} < 0.2$. Limits are also set on the production of scalar resonances, and model-independent cross section limits have been extracted as a function of diphoton invariant mass for any resonant $\gamma \gamma$ production process. These results extend the sensitivity of the previous search performed by the CMS experiment~\cite{Khachatryan:2016yec} and are compatible with those reported by the ATLAS Collaboration in Ref.~\cite{Aaboud:2017yyg}. In the large extra-dimensional model of Arkani-Hamed, Dimopoulos, and Dvali, exclusion limits on the string mass scale are set in the range $5.6 < \MS < 9.7\TeV$, depending on the specific model convention. These results extend the current best lower limits on \MS from the diphoton channel as presented in Ref.~\cite{Aaboud:2017yyg}. Additionally, the first exclusion limits are set in the two-dimensional parameter space of a continuum clockwork model.

\begin{acknowledgments}

We congratulate our colleagues in the CERN accelerator departments for the excellent performance of the LHC and thank the technical and administrative staffs at CERN and at other CMS institutes for their contributions to the success of the CMS effort. In addition, we gratefully acknowledge the computing centers and personnel of the Worldwide LHC Computing Grid for delivering so effectively the computing infrastructure essential to our analyses. Finally, we acknowledge the enduring support for the construction and operation of the LHC and the CMS detector provided by the following funding agencies: BMBWF and FWF (Austria); FNRS and FWO (Belgium); CNPq, CAPES, FAPERJ, FAPERGS, and FAPESP (Brazil); MES (Bulgaria); CERN; CAS, MoST, and NSFC (China); COLCIENCIAS (Colombia); MSES and CSF (Croatia); RPF (Cyprus); SENESCYT (Ecuador); MoER, ERC IUT, and ERDF (Estonia); Academy of Finland, MEC, and HIP (Finland); CEA and CNRS/IN2P3 (France); BMBF, DFG, and HGF (Germany); GSRT (Greece); NKFIA (Hungary); DAE and DST (India); IPM (Iran); SFI (Ireland); INFN (Italy); MSIP and NRF (Republic of Korea); MES (Latvia); LAS (Lithuania); MOE and UM (Malaysia); BUAP, CINVESTAV, CONACYT, LNS, SEP, and UASLP-FAI (Mexico); MOS (Montenegro); MBIE (New Zealand); PAEC (Pakistan); MSHE and NSC (Poland); FCT (Portugal); JINR (Dubna); MON, RosAtom, RAS, RFBR, and NRC KI (Russia); MESTD (Serbia); SEIDI, CPAN, PCTI, and FEDER (Spain); MOSTR (Sri Lanka); Swiss Funding Agencies (Switzerland); MST (Taipei); ThEPCenter, IPST, STAR, and NSTDA (Thailand); TUBITAK and TAEK (Turkey); NASU and SFFR (Ukraine); STFC (United Kingdom); DOE and NSF (USA).

\hyphenation{Rachada-pisek} Individuals have received support from the Marie-Curie program and the European Research Council and Horizon 2020 Grant, contract No. 675440 (European Union); the Leventis Foundation; the A. P. Sloan Foundation; the Alexander von Humboldt Foundation; the Belgian Federal Science Policy Office; the Fonds pour la Formation \`a la Recherche dans l'Industrie et dans l'Agriculture (FRIA-Belgium); the Agentschap voor Innovatie door Wetenschap en Technologie (IWT-Belgium); the F.R.S.-FNRS and FWO (Belgium) under the ``Excellence of Science - EOS" - be.h project n. 30820817; the Ministry of Education, Youth and Sports (MEYS) of the Czech Republic; the Lend\"ulet (``Momentum") Program and the J\'anos Bolyai Research Scholarship of the Hungarian Academy of Sciences, the New National Excellence Program \'UNKP, the NKFIA research grants 123842, 123959, 124845, 124850 and 125105 (Hungary); the Council of Science and Industrial Research, India; the HOMING PLUS program of the Foundation for Polish Science, cofinanced from European Union, Regional Development Fund, the Mobility Plus program of the Ministry of Science and Higher Education, the National Science Center (Poland), contracts Harmonia 2014/14/M/ST2/00428, Opus 2014/13/B/ST2/02543, 2014/15/B/ST2/03998, and 2015/19/B/ST2/02861, Sonata-bis 2012/07/E/ST2/01406; the National Priorities Research Program by Qatar National Research Fund; the Programa Estatal de Fomento de la Investigaci{\'o}n Cient{\'i}fica y T{\'e}cnica de Excelencia Mar\'{\i}a de Maeztu, grant MDM-2015-0509 and the Programa Severo Ochoa del Principado de Asturias; the Thalis and Aristeia programs cofinanced by EU-ESF and the Greek NSRF; the Rachadapisek Sompot Fund for Postdoctoral Fellowship, Chulalongkorn University and the Chulalongkorn Academic into Its 2nd Century Project Advancement Project (Thailand); the Welch Foundation, contract C-1845; and the Weston Havens Foundation (USA).

\end{acknowledgments}

\bibliography{auto_generated}
\cleardoublepage \appendix\section{The CMS Collaboration \label{app:collab}}\begin{sloppypar}\hyphenpenalty=5000\widowpenalty=500\clubpenalty=5000\input{EXO-17-017-authorlist.tex}\end{sloppypar}
\end{document}

%% file: EXO-17-017-authorlist.tex
\vskip\cmsinstskip
\textbf{Yerevan Physics Institute, Yerevan, Armenia}\\*[0pt]
A.M.~Sirunyan, A.~Tumasyan
\vskip\cmsinstskip
\textbf{Institut f\"{u}r Hochenergiephysik, Wien, Austria}\\*[0pt]
W.~Adam, F.~Ambrogi, E.~Asilar, T.~Bergauer, J.~Brandstetter, M.~Dragicevic, J.~Er\"{o}, A.~Escalante~Del~Valle, M.~Flechl, R.~Fr\"{u}hwirth\cmsAuthorMark{1}, V.M.~Ghete, J.~Hrubec, M.~Jeitler\cmsAuthorMark{1}, N.~Krammer, I.~Kr\"{a}tschmer, D.~Liko, T.~Madlener, I.~Mikulec, N.~Rad, H.~Rohringer, J.~Schieck\cmsAuthorMark{1}, R.~Sch\"{o}fbeck, M.~Spanring, D.~Spitzbart, A.~Taurok, W.~Waltenberger, J.~Wittmann, C.-E.~Wulz\cmsAuthorMark{1}, M.~Zarucki
% ...
\vskip\cmsinstskip
% ...
1:  Also at Vienna University of Technology, Vienna, Austria\\
% ...
